\documentclass[traditabstract,twocolumn,a4paper]{aa}
\usepackage{graphicx}
\usepackage{natbib}
\usepackage{amssymb}
\usepackage{hyperref}
\usepackage{txfonts}
\usepackage{rotating,bm}

\begin{document}

\title{
PORTA: A three-dimensional multilevel radiative transfer code for modeling the intensity and polarization of spectral lines with massively parallel computers
}

\titlerunning{3D Multilevel Radiative Transfer with Polarization}

\author{
Ji\v{r}\'{\i} \v{S}t\v{e}p\'an\inst{1}
\and
Javier Trujillo Bueno\inst{2,3,4}
}

\institute{
Astronomical Institute ASCR, v.v.i., Ond\v{r}ejov, Czech Republic.
\email{jiri.stepan@asu.cas.cz}
\and
Instituto de Astrof\'{\i}sica de Canarias, V\'{\i}a L\'actea s/n, E-38205 La Laguna, Tenerife, Spain.
\email{jtb@iac.es}
\and
Departamento de Astrof\'{\i}sica, Universidad de La Laguna (ULL), E-38206 La Laguna, Tenerife, Spain
\and
Consejo Superior de Investigaciones Cient\'{\i}ficas, Spain
}

\date{Received XXXX; accepted XXXX}

\abstract{
The interpretation of the intensity and polarization of the spectral line radiation produced in the atmosphere of the Sun and of other stars requires solving a radiative transfer problem that can be very complex, especially when the main interest lies in modeling the spectral line polarization produced by scattering processes and the Hanle and Zeeman effects. One of the difficulties is that the plasma of a stellar atmosphere can be highly inhomogeneous and dynamic, which implies the need to solve the non-equilibrium problem of the generation and transfer of polarized radiation in realistic three-dimensional (3D) stellar atmospheric models. Here we present PORTA, an efficient multilevel radiative transfer code we have developed for the simulation of the spectral line polarization caused by scattering processes and the Hanle and Zeeman effects in 3D models of stellar atmospheres. The numerical method of solution is based on the non-linear multigrid iterative method and on a novel short-characteristics formal solver of the Stokes-vector transfer equation which uses monotonic B\'ezier interpolation. Therefore, with PORTA the computing time needed to obtain at each spatial grid point the self-consistent values of the atomic density matrix (which quantifies the excitation state of the atomic system) scales linearly with the total number of grid points. Another crucial feature of PORTA is its parallelization strategy, which allows us to speed up the numerical solution of complicated 3D problems by several orders of magnitude with respect to sequential radiative transfer approaches, given its excellent linear scaling with the number of available processors. The PORTA code can also be conveniently applied to solve the simpler 3D radiative transfer problem of unpolarized radiation in multilevel systems.
}


\maketitle

\section{Introduction\label{sec:intro}}

This paper describes a computer program we have developed for solving, in three-dimensional (3D) models of stellar atmospheres, the problem of the generation and transfer of spectral line polarization taking into account anisotropic radiation pumping and the Hanle and Zeeman effects in multilevel systems. The numerical method of solution is based on a highly convergent iterative method, whose convergence rate is insensitive to the grid size, and on an accurate short-characteristics formal solver of the Stokes-vector transfer equation that uses monotonic B\'ezier interpolation. A key feature of our multilevel code called PORTA (POlarized Radiative TrAnsfer) is its parallelization strategy, which allows us to speed up the numerical solution of complicated 3D problems by several orders of magnitude with respect to sequential radiative transfer approaches. 

The multilevel radiative transfer problem currently solved by PORTA is the so-called non-LTE problem of the 2nd kind (\citet{ll04}, hereafter LL04; see also \citet{jtb09}), where the phenomenon of scattering in a spectral line is described as the temporal succession of statistically-independent events of absorption and re-emission (complete frequency redistribution, or CRD). This is a formidable numerical problem that implies calculating, at each spatial grid point of the (generally magnetized) 3D stellar atmosphere model under consideration, the values of the multipolar components of the atomic density matrix corresponding to each atomic level of total angular momentum $J$. These $\rho^K_Q(J)$ elements, with $K=0, \dots, 2J$ and $Q=-K, \dots, K$, quantify the overall population of each level $J$ ($\rho^0_0(J)$), the population imbalances between its magnetic sublevels ($\rho^K_0(J)$, with $K>0$), and the quantum coherence between each pair of them ($\rho^K_Q(J)$, with $K>0$ and $Q{\ne}0$). The values of these density-matrix elements have to be consistent with the intensity, polarization, and symmetry properties of the incident radiation field generated within the medium. Finding these density-matrix values requires solving jointly the radiative transfer (RT) equations for the Stokes parameters ($\vec I(\nu,\vec\Omega)=(I,Q,U,V)^{\rm T}$, with $\nu$ and $\vec\Omega$ the frequency and direction of propagation of the radiation beam under consideration) and the statistical equilibrium equations (SEE) for the $\rho^K_Q(J)$ elements. These $\rho^K_Q(J)$ elements, at each spatial grid point of the 3D atmospheric model and for each level $J$ of the considered atomic model, provide a complete description of the excitation of each level $J$. As a result, the radiative transfer coefficients (i.e., the emission vector and the propagation matrix of the Stokes-vector transfer equation) corresponding to each line transition depend on the $\rho^K_Q(J)$ values of the upper ($J=J_u$) and lower ($J=J_l$) line levels. Once the self-consistent $\rho^K_Q(J)$ values are obtained at each point within the medium, PORTA solves the Stokes-vector transfer equation to obtain the emergent Stokes profiles for any desired line transition and line of sight. Obviously, this last step is computationally cheap compared with the time needed to obtain the self-consistent solution for the $\rho^K_Q(J)$ elements.

When polarization phenomena are neglected the only non-zero density-matrix element is $\rho^0_0(J)$ (proportional to the overall population of each $J$ level) and the only non-zero Stokes parameter is the specific intensity $I(\nu,\vec\Omega)$ for each of the radiative transitions in the model atom under consideration. This non-LTE problem of the 1st kind \citep[e.g.,][]{mihalas78} is a particular case of the above-mentioned non-LTE problem of the 2nd kind. In other words, the 3D multilevel radiative transfer code described here can also be applied to solve the standard non-LTE multilevel problem on which much of today's quantitative stellar spectroscopy is based. For this reason, PORTA provides options for spectropolarimetry and for spectroscopy (see Appendix\,\ref{app:bench}).

An overview on 3D radiative transfer codes for unpolarized radiation can be found in \citet{carlsson08}. Information on numerical methods for the transfer of spectral line polarization can be found in some reviews \citep[e.g.,][]{jtb03,nag09} and research papers \citep[e.g.,][]{rees89,pal98,rms99,rms03,rms11,sampoorna10,anusha11}. To the best of our knowledge this is the first time that a computer program suitable for massively parallel computers has been developed to solve the multilevel problem of the generation and transfer of spectral line polarization resulting from scattering processes and the Hanle and Zeeman effects in 3D stellar atmosphere models. The problem of the generation and transfer of spectral line polarization with partial frequency redistribution (PRD) is being increasingly considered in the literature \citep[e.g.,][]{sampoorna10b,bel12}, but assuming relatively simple model atoms suitable only for some resonance lines.

After presenting in Sect.\,\ref{sec:formul} the formulation of the RT problem, in Sect.\,\ref{sec:besser} we explain our formal solver of the 3D Stokes-vector transfer equation, which is based on Auer's (2003) suggestion of monotonic B\'ezier interpolation within the framework of the short-characteristics approach. An additional important point is the parallelization strategy we have developed for taking advantage of massively parallel computers, which we detail in Sect.\,\ref{sec:snake} following our explanation of the formal solver. The iterative method we have implemented, explained in Sect.\,\ref{sec:mg}, is based on the non-linear multigrid method for radiative transfer applications proposed by \citet{pfb97}. We present useful benchmarks and comparisons of the multigrid iterative option of our code with another option based on the Jacobian iterative method on which one-dimensional (1D) multilevel codes for solving the non-LTE problem of the 2nd kind are based \citep[e.g.,][]{rms03,stepan11}. Finally, in Sect.\,\ref{sec:con} we present our conclusions with a view to future research.

We have already applied PORTA to investigate the intensity and linear polarization of some strong chromospheric lines in a model of the extended solar atmosphere resulting from state-of-the-art 3D magneto-hydrodynamic (MHD) simulations (e.g., \v{S}t\v{e}p\'an et al. 2012). However, for the benchmarks presented in this paper, whose aim is a detailed description of PORTA, we have found it more suitable to choose the 3D model atmosphere and the five-level atomic model detailed in Appendix~\ref{app:atom}. The software and hardware tools are summarized in Appendix~\ref{app:bench}.

\section{The radiative transfer problem\label{sec:formul}}

The multilevel non-LTE problem considered here for the generation and transfer of spectral line polarization is that outlined in Sect.\,3 of \citet{stepan11}, where we assumed one-dimensional (1D), plane-parallel atmospheric models \citep[see also][]{rms03}, while the aim here is to describe the computer program we have developed for solving in Cartesian coordinates the same multilevel radiative transfer problem but in three-dimensional (3D) stellar atmospheric models. As shown below, the development of a robust multilevel 3D code is not simply an incremental step with respect to the 1D case, given the need to develop and implement an accurate 3D formal solver, a highly convergent iterative scheme based on multiple grids, and a suitable parallelization strategy to take advantage of today's massively parallel computers.

A detailed presentation of all the physics and relevant equations necessary to understand the radiation transfer problem solved in this paper can be found in Chapter\,7 of LL04. Our aim here is to solve jointly the Stokes-vector transfer equation (corresponding to each radiative transition in the model atom under consideration) and the statistical equilibrium and conservation equations for the multipolar components of the atomic density matrix (corresponding to each level $J$). We take into account the possibility of quantum coherence (or interference) between pairs of magnetic sublevels pertaining to any given $J$~level, but neglect quantum interference between sublevels pertaining to different $J$~levels. Neglecting $J$-state interference is a very suitable approximation for modeling the line-core polarization, which is where the Hanle effect in most solar spectral lines operates \citep[see][]{bel11}. In the absence of $J$-state interference, the general number of $\rho^K_Q(J)$ unknowns for each level $J$ is $(2J+1)^2$, at each spatial grid point. We note that in the unpolarized case there is only one unknown associated to each $J$~level (i.e., $\rho^0_0(J)$).

In this paper we focus on the multilevel model atom (see Sects.\,7.3 and 7.13c of LL04), in which quantum interference between sublevels pertaining to different $J$~levels are neglected. However, it is important to note that the same iterative method, formal solver, and the overall logical structure of our code PORTA are very suitable for solving the same type of problem but considering other model atoms and/or magnetic field regimes (see Chapter 7 of LL04).

The emission vector and the propagation matrix of the Stokes-vector transfer equation depend on the local values of the $\rho^K_Q$ elements  (with $K=0,\dots,2J$ and $Q=-K, \dots, K$) of the upper ($i$) and lower ($j$) line levels (see Sect.\,7.2.b in LL04). Given an estimation of these $\rho^K_Q(J)$ elements for each $J$~level at all spatial grid points, the formal solution of the Stokes-vector transfer equation for each radiative transition allows us to obtain the ensuing Stokes parameters at each spatial point within the medium, for each discretized frequency and ray direction. After angle and frequency integration one can obtain the radiation field tensors
\begin{equation}
J^K_Q(ij)=\sum_{p=0}^3\oint\frac{{\rm d}\vec\Omega}{4\pi}\mathcal{T}^K_Q(p,\vec\Omega)\int {\rm d}\nu\, I_p(\nu,\vec\Omega)\phi_{ij}(\nu-\nu_{ij})\,,
\end{equation}
where $\phi_{ij}$ is the line absorption profile and $\mathcal{T}^K_Q(p,\vec \Omega)$ the spherical irreducible tensors given in Table 5.6 of LL04, and $(I_0,I_1,I_2,I_3)^{\rm T}\equiv(I,Q,U,V)^{\rm T}$ is the so-called Stokes vector. These radiation field tensors, defined for $K=0,1,2$ and $Q=-K, \dots, K$, specify the symmetry properties of the radiation field that illuminates each point within the medium. These quantities are of fundamental importance because they determine the radiative rates that enter the statistical equilibrium equations (see Sect.\,7.2.a in LL04). After the discretization of the spatial dependence these equations can be expressed as
\begin{equation}
{\bm{\mathcal{L}}_l}\,{\vec {\rho}}_l\,=\,{\vec f}_l\,,   \label{eq:nonlte}
\end{equation}
where $\bm{\mathcal{L}}_l$ is a block-diagonal matrix formed by $\mathcal {N}_l$ submatrices, $\mathcal {N}_l$ being the number of points of the spatial grid of resolution level $l$ (the larger the positive integer number $l$ the finer the grid). ${\rm NL}\times{\rm NL}$ is the size of each submatrix, ${\rm NL}$ being the total number of $\rho^K_Q$ unknowns at each spatial grid point. The length of the vector $\vec\rho_l$ of $\rho^K_Q$ unknowns and of the known vector ${\vec f}_l$ is ${\rm NL}{\times}{{\mathcal {N}}}_l$. The coefficients of the block-diagonal matrix $\bm{\mathcal{L}}_{l}$ depend on the collisional rates, which depend on the local values of the thermodynamical variables, and on the radiative rates, which depend on the radiation field tensors $J^K_Q(ij)$, whose computation requires solving the Stokes-vector transfer equation for each radiative transition $i{\rightarrow}j$. Since the radiative transfer coefficients depend on the $\rho^K_Q$ unknowns the problem is non-linear, in addition to non-local.

To solve this type of problem we need a fast and accurate formal solver of the Stokes-vector transfer equation and a suitable iterative method capable of finding rapidly the density matrix elements $\vec\rho_l$ these that Eq.~(\ref{eq:nonlte}) is satisfied when the radiation field tensors, which appear in the block-diagonal matrix $\bm{\mathcal{L}}_{l}$, are calculated from such $\vec\rho_l$ elements via the solution of the Stokes-vector transfer equation. The 1D multilevel code described in Appendix\,A of \citet{stepan11} is based on the DELOPAR formal solver proposed by \citet{jtb03} and on a Jacobian iterative scheme, similar to that applied by \citet{rms03}, but generalized to the case of overlapping transitions. 

We turn now to explain the new formal solver we have developed for 3D Cartesian grids that is based on monotonic B\'ezier interpolation.

\section{BESSER: Monotonic B\'ezier formal solver of the Stokes-vector transfer equation\label{sec:besser}}

The transfer equation for the Stokes vector $\vec I=(I,Q,U,V)^{\rm T}$ can be written \citep[e.g.,][]{rees89,jtb03}
\begin{equation}
\frac{{\rm d}}{{\rm d} \tau}\vec I=\vec I\,-\,\vec{S}_{\rm eff},
\label{eq:rte}
\end{equation}
where $\tau$ is the optical distance along the ray under consideration 
(${\rm d}{\tau}=-\eta_I\,{\rm d} s$, with $s$ the geometrical distance along the ray and $\eta_I$ the diagonal element of the $4\times 4$ propagation matrix $\vec K$), $\vec{S}_{\rm eff}=\vec{S}-\vec{K}'\vec{I}$ being $\vec{K}'=\vec{K}/{\eta_I}-\vec{1}$ (where $\vec{1}$ is the unit matrix and $\vec{S}=\vec{\epsilon}/\eta_I$, with $\vec\epsilon=(\epsilon_I,\epsilon_Q,\epsilon_U,\epsilon_V)^{\rm T}$ the emission vector resulting from spontaneous emission events). The formal solution of this equation is

\begin{equation}
\vec I_{\rm O}=\vec I_{\rm M} {\rm e}^{-\tau_{\rm MO}}+\int_0^{\tau_{\rm MO}} {\rm d} t\, \left[\vec S(t)-\vec K'(t)\vec I(t)\right]{\rm e}^{-t}\;,
\label{eq:fsint}
\end{equation}
where the ray or pencil of radiation of frequency $\nu$ propagates along the direction $\vec\Omega$, from the upwind point ${\rm M}$ (where the Stokes vector $\vec I_{\rm M}$ is assumed to be known) towards the spatial point ${\rm O}$ of interest (where the Stokes vector $\vec I_{\rm O}$ is sought), and $t$ is the optical path along the ray (measured from ${\rm O}$ to ${\rm M}$; see Fig.~\ref{fig:stencil}).

The numerical solution of Eq.~(\ref{eq:fsint}) allows us to obtain, from the current estimates of the emission vector $\vec\epsilon$ and propagation matrix $\vec{K}$, the Stokes parameters at each spatial grid point ${\rm O}$ within the 3D medium, for all discretized radiation frequencies and directions. We note that the unpolarized version of Eq.~(\ref{eq:fsint}) can be easily obtained by taking $\vec I_{\rm O}{\to}I_{\rm O}$, $\vec I_{\rm M}{\to}I_{\rm M}$, $\vec S(t){\to}S(t)$, and $\vec K'(t){\to}0$.

\subsection{The short characteristics method}

\begin{figure}
\begin{center}
\includegraphics[width=0.8\columnwidth]{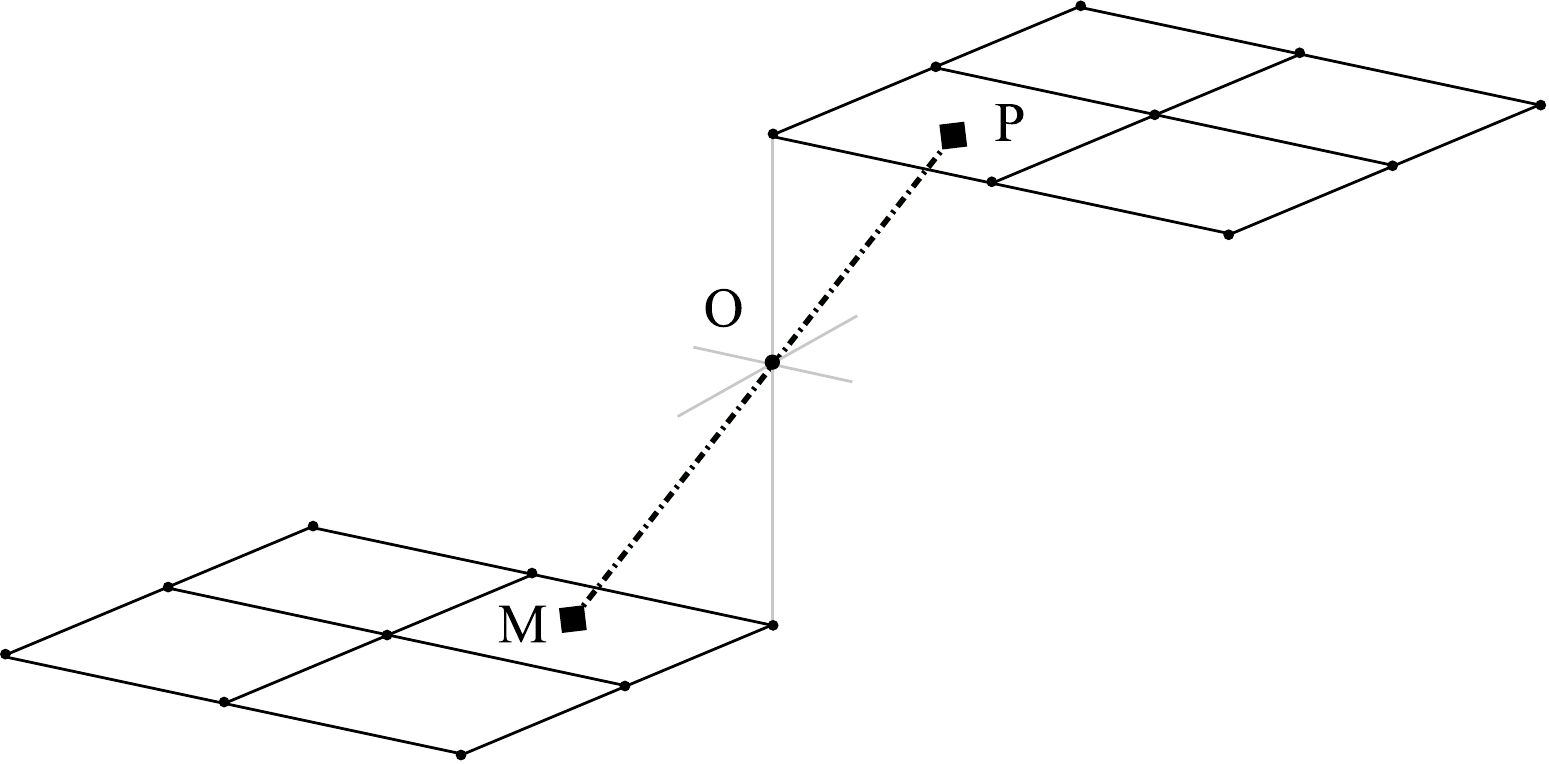}
\end{center}
\caption{Short-characteristics in a three-dimensional Cartesian rectilinear grid.}
\label{fig:stencil}
\end{figure}

The short-characteristics (SC) method was proposed by \citet{ka88} 
to solve the unpolarized version of Eq.~(\ref{eq:fsint}) for the specific intensity \citep[see also][]{ap94,auer94,pfb99}. Consider three consecutive spatial points M, O, and P along the ray under consideration, with M the upwind point, P the downwind point, and O the point where the Stokes~$I$ parameter is being sought, for the given frequency and ray direction (see Fig.~\ref{fig:stencil}). The aim of the original SC method is to solve the unpolarized version of Eq.~(\ref{eq:fsint}) along the MO segment in order to compute the specific intensity $I(\nu,\vec{\Omega})$. The original SC method is based on the approximation of parabolic interpolation of the source function $S(t)$ between the upwind point M of the short characteristics, the grid point O, and the downwind point P (see Fig.~\ref{fig:stencil}). In 2D and 3D grids, the upwind and downwind points of the SC do not generally coincide with any spatial grid node and the radiation transfer quantities (i.e., the emission and absorption coefficients) have to be interpolated from the nearby 9-point (biquadratic case) or 4-point (bilinear case) stencils of the discrete grid points. In the unpolarized and polarized options of PORTA both biquadratic and bilinear interpolation are implemented. Bilinear interpolation is sufficient in the fine grids of today's MHD models. The upwind specific intensity or the Stokes vector $\vec I_{\rm M}$ need to be interpolated from the same grid nodes. Proper topological ordering of the grid points is therefore necessary for every direction of the short characteristics. We note that the intersection points M and P may be located on a vertical plane of the grid instead of a horizontal one.

\begin{figure}
\begin{center}
\includegraphics[width=0.8\columnwidth]{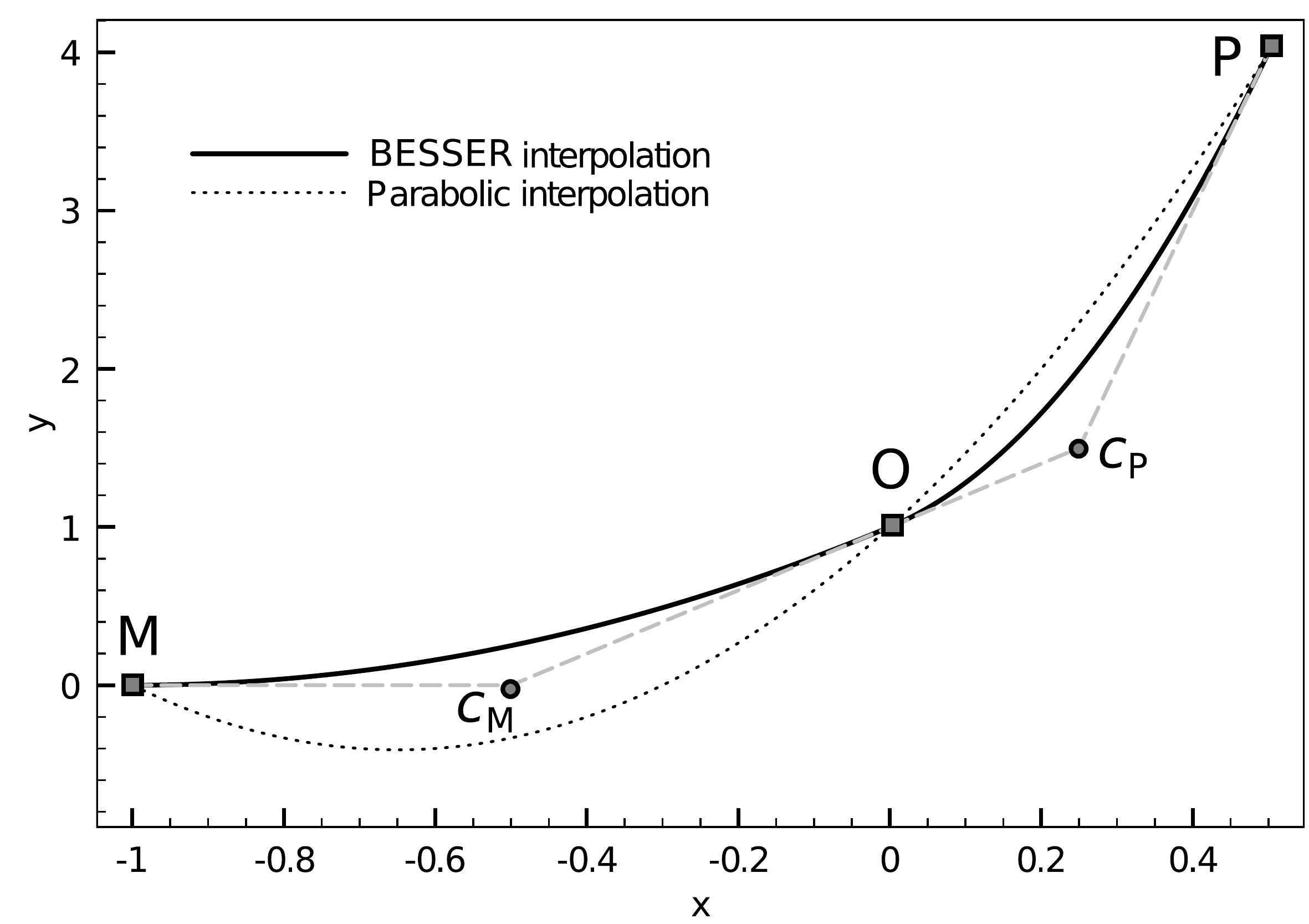}
\end{center}
\caption{
Parabolic and BESSER interpolation using the three successive points M, O, and P. Dotted line: parabolic interpolation may create a spurious extremum between points M and O. Solid line: interpolation using our BESSER method with continuous derivative at point O. The control points of the intervals, whose $y$-coordinates are denoted by $c_{\rm M}$ and $c_{\rm P}$, define tangents to the B\'ezier splines in their endpoints. The $x$-coordinates of the control points are located at the center of the corresponding intervals.
}
\label{fig:bezi}
\end{figure}

The DELO method proposed by \citet{rees89} can be considered a possible generalization of the scalar SC method to the radiative transfer problem of polarized radiation. That formal solver of the Stokes-vector transfer equation is, however, based on linear interpolation of the source function ${\vec S}_{\rm eff}$ between points M and O. \citet{jtb03} demonstrated that significantly more accurate solutions can be obtained by using instead a formal solver he calls DELOPAR, which is based on the choice in Eq.~(\ref{eq:fsint}) of parabolic interpolation for $\vec S$ (between points M, O, and P) and linear interpolation for $\vec{K}'\vec{I}$ (between points M and O). He showed that with DELOPAR the accuracy of the self-consistent solution rapidly improves as the spatial resolution level $l$ of the spatial grid is increased. The first version of the computer program NICOLE of \citet{hsn00}, for the synthesis and inversion of Stokes profiles induced by the Zeeman effect, used DELOPAR as the formal solver.

In smooth and/or suitably discretized model atmospheres, the DELOPAR method provides accurate results. However, in the presence of abrupt changes in the physical properties of the atmospheric model, the parabolic interpolation suffers from non-monotonic interpolation between otherwise monotonic sequences of discrete points. Such spurious extrema of the interpolant decrease the accuracy of the solution and can also lead to unrealistic or even unphysical Stokes parameters at the grid point O under consideration (see the dotted line in Fig.~\ref{fig:bezi}). In addition, the parabolic interpolation may occasionally lead to the divergence of the whole numerical solution.

To overcome these difficulties, \citet{auer03} suggested an interpolation based on the use of monotonic B\'ezier splines. Some formal solvers based on this idea have already been implemented \citep{koe08,hayek08,stepan-3d12,hol12,jaime13}. In this section, we describe in detail the accurate formal solver we have developed, pointing out a significant difference with the original proposal of \citet{auer03}. We call it BESSER (BEzier Spline SolvER).

\subsection{Monotonic spline interpolation with continuous first derivative
\label{ssec:deloparalg}}

Following \citet{auer03}, our BESSER algorithm is based on the use of piecewise monotonic quadratic B\'ezier splines. The control points of the splines can be used to preserve monotonicity of the interpolant, because the interpolant is contained in an envelope defined by the tangents of the spline in its endpoints and of the control point which is located in the intersection of these tangents (see Fig.~\ref{fig:bezi}). As shown below, we achieve a smooth connection of the B\'ezier splines in the central point O by imposing a continuous first derivative of the interpolant. This improvement over the original treatment of \citet{auer03} leads to a symmetrical interpolation independently of the choice of the interpolation direction (MOP for one direction of the ray propagation or POM for the opposite direction of the ray). An additional attractive feature is that our BESSER method always provides reliable values for the diagonal of the $\Lambda$-operator, i.e., in the interval $[0,1)$, used in methods based on the Jacobi iteration.

\begin{figure}
\begin{center}
\includegraphics[width=0.8\columnwidth]{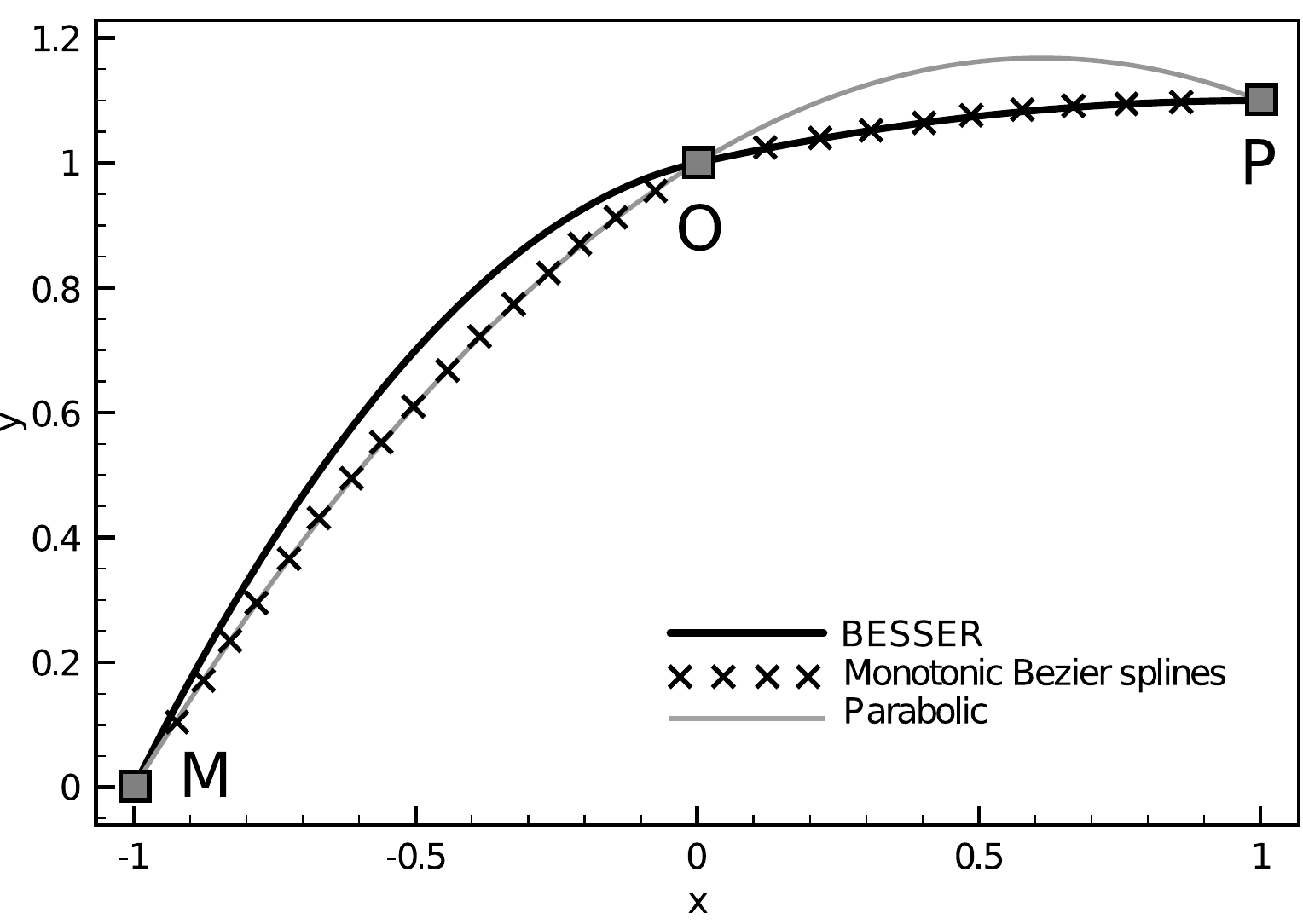}
\end{center}
\caption{
The treatment of an overshoot in the downwind interval OP by three different methods. Black solid line: BESSER  implementation with continuous derivative at point O and the $c_{\rm P}$ overshoot correction of the control point. Crosses: piecewise monotonic quadratic B\'ezier spline interpolation. Solid gray line: parabolic interpolation. We note that the piecewise monotonic quadratic B\'ezier interpolation coincides with the parabolic interpolation in the MO segment because the overshoot in the OP interval does not affect the upwind interpolation between M and O.
}
\label{fig:opovershoot}
\end{figure}

Given a quantity $y$ (e.g., the source function) defined at three successive points $x_{\rm M}$, $x_{\rm O}$, and $x_{\rm P}$, we use two quadratic B\'ezier splines to interpolate $y$ between points M and O and between points O and P (see Fig.~\ref{fig:bezi}). First, we look for an optimal interpolation in the interval MO. For the sake of simplicity, we parametrize the $x$-coordinate in this interval by a dimensionless parameter $u=(x-x_{\rm M})/h_{\rm M}$, where $h_{\rm M}=x_{\rm O}-x_{\rm M}$. The B\'ezier spline in the interval MO is a parabola passing through points M and O. The  derivatives at such points are defined by the position of the control point whose $y$-coordinate is $c_{\rm M}$ (see Fig.~\ref{fig:bezi}). The equation for such a spline reads \citep{auer03}
\begin{equation}
y(u)=(1-u)^2y_{\rm M} + 2u(1-u)c_{\rm M} + u^2y_{\rm O}\,,\qquad u\in[0,1]\;.
\label{eq:beziers}
\end{equation}
Similarly, one can define a B\'ezier spline between points O and P by doing the formal changes $y_{\rm M}\to y_{\rm O}$, $y_{\rm O}\to y_{\rm P}$, and $u =(x-x_{\rm O})/h_{\rm P}$, where $h_{\rm P}=x_{\rm P}-x_{\rm O}$; the $y$-coordinate of the ensuing control point is denoted by $c_{\rm P}$ (see Fig.~\ref{fig:bezi}).

We look for the values of $c_{\rm M}$ and $c_{\rm P}$ that satisfy the following conditions: (1) if the sequence $y_{\rm M}$, $y_{\rm O}$, $y_{\rm P}$ is monotonic, then the interpolation is monotonic in the whole interval $[x_{\rm M},x_{\rm P}]$; (2) if the sequence of $y_i$ values is not monotonic, then the interpolant has the only local extremum at O; and (3) the first derivative of the interpolant at point O should be continuous. The ensuing algorithm proceeds as follows:
\begin{enumerate}
\item Calculate the quantities $d_{\rm M}=(y_{\rm O}-y_{\rm M})/h_{\rm M}$, $d_{\rm P}=(y_{\rm P}-y_{\rm O})/h_{\rm P}$.
\item If the sequence $y_{\rm M}$, $y_{\rm O}$, $y_{\rm P}$ is not monotonic (i.e., if $d_{\rm M}d_{\rm P}\le 0$), then set $c_{\rm M}=c_{\rm P}=y_{\rm O}$ and exit the algorithm. The derivative of the splines at point O is equal to zero, leading to a local extremum at the central point.
\item Estimate the derivative at point O,
\begin{equation}
y'_{\rm O}=\frac{h_{\rm M}d_{\rm P}+h_{\rm P}d_{\rm M}}{h_{\rm M}+h_{\rm P}}
\label{eq:deriv}
\end{equation}
\citep[see Eq. 7 of][and references therein]{auer03}. This derivative is equal to that provided at the same point by parabolic interpolation among points M, O, and P. Moreover, in contrast with Eq. (12) of \citet{auer03}, it is an expression that relates $y'$ (e.g., the source function derivative) linearly with the $y$-values (e.g., with the source function values).
\item Calculate the initial positions of the control points, $c_{\rm M}=y_{\rm O}-\frac{h_{\rm M}}{2}y'_{\rm O}$, and $c_{\rm P}=y_{\rm O}+\frac{h_{\rm P}}{2}y'_{\rm O}$.\footnote{The control points calculated this way lead to a unique parabolic interpolation among points MOP. If the algorithm is stopped here, the resulting formal solver would be equivalent to the standard parabolic interpolation.}
\item Check that $\min(y_{\rm M},y_{\rm O})\le c_{\rm M} \le \max(y_{\rm M},y_{\rm O})$. If the condition is satisfied, then go to step 7, otherwise continue with step 6.
\item If the condition in step 5 is not satisfied, then there is an overshoot of the interpolant in the interval MO. Set $c_{\rm M}=y_{\rm M}$, so that the first derivative at M is equal to zero and the overshoot is corrected. Since the value of $c_{\rm P}$ is not of interest for the formal solution between M and O, exit the algorithm.
\item Check if $\min(y_{\rm O},y_{\rm P})\le c_{\rm P} \le \max(y_{\rm O},y_{\rm P})$. If this condition is not satisfied, then set $c_{\rm P}=y_{\rm P}$ so that the overshoot in the interval OP is corrected.
\item Calculate the new derivative at O, $y'_{\rm O} = (c_{\rm P}-y_{\rm O})/(h_{\rm P}/2)$, using the  corrected value of $c_{\rm P}$ calculated in step 7.
\item Calculate a new $c_{\rm M}$ value to keep the derivative at O smooth. It is easy to realize that this change cannot produce an overshoot in the MO interval, hence the solution remains monotonic with a continuous derivative.
\end{enumerate}
 
Steps 8 and 9 of the above-mentioned algorithm, dealing with correction of the overshoots in the downwind interval followed by modification of the $c_{\rm M}$ upwind control point value, are not part of the original algorithm of \citet{auer03} in which the derivative at point O can be discontinuous. We have found that it is suitable to guarantee the smoothness of the derivative, and this can be done with only a small increase in the computing time with respect to the DELOPAR method. Our BESSER interpolation is stable; that is, the interpolant varies smoothly with smooth changes of the M, O, and P points. No abrupt changes of the splines occur that could negatively affect the stability of the iterative method.

In contrast to some other formal solvers based on the idea of quadratic B\'ezier splines \citep[e.g., one of the two Bezier methods discussed by][]{jaime13}, our BESSER algorithm guarantees that a monotonic sequence of the MOP points leads to a monotonic interpolant in all situations. This fact is of critical importance in 2D and 3D grids in which $\tau_{\rm MO}$ and $\tau_{\rm OP}$ may differ significantly because of unpredictable intersections of the grid planes, especially if periodic boundary conditions are considered. Such large differences often lead to overshoots, unless treated by BESSER or a similarly suitable strategy. Formal solvers based on cubic Bezier splines \citep[e.g.,][]{jaime13} could be developed to preserve the continuity of $y'_{\rm O}$, but they may fail to accurately interpolate quadratic functions even in fine grids when using Auer's (2003) Eq. (12) for $y'_{\rm O}$ (see Sect.\,3.4). 

An alternative formal solver, which uses cubic Hermite splines, has been presented by \citet{ibgui13}. However, the way of fixing the derivatives at the end points M and P of the SC to the values corresponding to the linear interpolation case may cause loss of accuracy of the formal solver.

\subsection{Formal solution of the vectorial radiative transfer equation with BESSER}

\begin{table*}
\caption{\label{tab:expand}Taylor expansion of the $\omega_{\rm M,O,P}$ coefficients for small optical path intervals}
\centering
\begin{tabular}{ccl}
\hline\hline
Coefficient & ${\rm max}\;t$ & Expansion \\
\hline
$\omega_{\rm M}$ & 0.14 & $(t(t(t(t(t(t((140-18t)t-945)+5400)-25200)+90720)-226800)+302400))/907200$\\
$\omega_{\rm O}$ & 0.18 & $(t(t(t(t(t(t((10-t)t-90)+720)-5040)+30240)-151200)+604800))/1814400$\\
$\omega_{\rm P}$ & 0.18 & $(t(t(t(t(t(t((35-4t)t-270)+1800)-10080)+45360)-151200)+302400))/907200$\\
\hline
\end{tabular}
\tablefoot{Taylor expansion of the interpolation coefficients of the BESSER method for small $\tau_{\rm MO}$ values (for the sake of notational simplicity we use $t\equiv\tau_{\rm MO}$). Column~2 of the table indicates the approximate maximum value of $\tau_{\rm MO}$ for which this expansion is more accurate, using double precision arithmetics, than the expressions given by Eqs.~(\ref{eq:omegam}--\ref{eq:omegap}). We use the Horner rule in Col.~3, which provides a better numerical accuracy and also reduces the number of multiplications in comparison to the explicit expansion of the Taylor power series.}
\end{table*}

The application of the B\'ezier interpolation for calculating the formal solution of Eq.~(\ref{eq:fsint}) proceeds as follows. We assume that the Stokes components of the vectorial source function $\vec S(t)$ vary, between points M and O, according to Eq.~(\ref{eq:beziers}) with the control points calculated using the BESSER algorithm described in the previous section. The term $\vec K'(t)\vec I(t)$ is assumed to change linearly in the same interval, as in the DELOPAR method. The integral in Eq. (\ref{eq:fsint}) can then be evaluated analytically and the Stokes parameters at point O can be expressed in the form \citep[see][for details of an analogous derivation using  parabolic interpolation of $\vec S$]{jtb03}
\begin{equation}
\vec\kappa^{-1}\vec I_{\rm O}=\left[e^{-\tau_{\rm MO}}-\psi_{\rm M}'\vec K_{\rm M}' \right]\vec I_{\rm M} +\omega_{\rm M}\vec S_{\rm M}+\omega_{\rm O}\vec S_{\rm O}+\omega_{\rm C}\vec c_{\rm M}\;,
\label{eq:fsbez}
\end{equation}
where
\begin{equation}
\vec\kappa^{-1}=\vec 1+\psi_{\rm O}'\vec K_{\rm O}'     \label{eq:kappa}
\end{equation}
is a $4{\times}4$ matrix and $\vec 1$ is the unit matrix. Multiplying Eq.~(\ref{eq:fsbez}) by $\vec\kappa$ gives the desired vector of Stokes parameters $\vec I_{\rm O}$ at point O. The coefficients $\psi_{\rm M}'$ and $\psi_{\rm O}'$ are the usual coefficients resulting from linear interpolation,
\begin{eqnarray}
\psi_{\rm M}'&=& \frac{1-e^{-\tau_{\rm MO}}(1+\tau_{\rm MO})}{\tau_{\rm MO}}\,,\\
\psi_{\rm O}'&=& \frac{e^{-\tau_{\rm MO}}+\tau_{\rm MO}-1}{\tau_{\rm MO}}\,.
\end{eqnarray}
Using the substitutions $h_{\rm M}=\tau_{\rm MO}$ and $u=1-t/\tau_{\rm MO}$ in Eq.~(\ref{eq:beziers}), one obtains for the $\omega_i$ coefficients the expressions
\begin{eqnarray}
\omega_{\rm M} &=& \frac{2-e^{-\tau_{\rm MO}}(\tau_{\rm MO}^2+2\tau_{\rm MO}+2)}{\tau_{\rm MO}^2}   \label{eq:omegam}\;,  \\
\omega_{\rm O} &=& 1-2\frac{e^{-\tau_{\rm MO}}+\tau_{\rm MO}-1}{\tau_{\rm MO}^2}\;,   \label{eq:omegao} \\
\omega_{\rm C} &=& 2\frac{\tau_{\rm MO}-2+e^{-\tau_{\rm MO}}(\tau_{\rm MO}+2)}{\tau_{\rm MO}^2}\;.\label{eq:omegap}
\end{eqnarray}

It is important to note that the accuracy of these expressions decreases as $\tau_{\rm MO}\ll 1$, due to the limited precision of the floating point computer arithmetics. Therefore, for small upwind optical paths we use instead the Taylor expansion of such expressions calculated at $\tau_{\rm MO}=0$ (see Table~\ref{tab:expand}).

An important quantity used in Jacobi-based iterative methods for the solution of non-LTE problems is the diagonal of the monochromatic $\Lambda$ operator at the point O under consideration, $\Lambda^*_{\vec\Omega\nu}$. It can be easily calculated from Eq.~(\ref{eq:fsbez}) for $I_{\rm O}$, by setting $\vec I_{\rm M}=\vec S_{\rm M}=\vec S_{\rm P}=(0,0,0,0)^{\rm T}$ and $\vec S_{\rm O}=(1,0,0,0)^{\rm T}$. It follows that $\Lambda^*_{\vec\Omega\nu}=\omega_{\rm O}+\omega_{\rm C} c_{\rm M}$. Given that in this case the source function has a local maximum at point O, we have $c_{\rm M}=1$, and we finally arrive at
\begin{equation}
\Lambda^*_{\vec\Omega\nu}=\omega_{\rm O}+\omega_{\rm C}=1+2\frac{e^{-\tau_{\rm MO}}(1+\tau_{\rm MO})-1}{\tau_{\rm MO}^2}\;.
\end{equation}
In contrast to the familiar parabolic solvers, no information about the interpolation coefficients in the preceding point is needed to determine the diagonal of the $\Lambda$ operator at the point O under consideration. It is easy to show that $\Lambda^*_{\vec\Omega\nu}\in [0,1)$, which is an important condition for the stability of the iterative method used to solve any non-LTE problem. This is particularly important for solving three-dimensional problems in which the upwind point M does not generally coincide with any grid node.

We have found by numerical experimentation that the time needed to perform one Jacobi iteration using our BESSER formal solver is only ${\lesssim}1\%$ slower than when using instead DELOPAR, because of the need to determine the $c_{\rm M}$ and $c_{\rm P}$ values of the control points following the algorithm described in Sect.\,\ref{ssec:deloparalg}. The computation of $\vec\kappa$ and $\vec\kappa^{-1}$ (see Eq.~\ref{eq:kappa}), the calculation of the transfer coefficients, and their interpolation in the upwind and downwind points takes most of the computing time per iterative step.

\begin{figure}
\begin{center}
\includegraphics{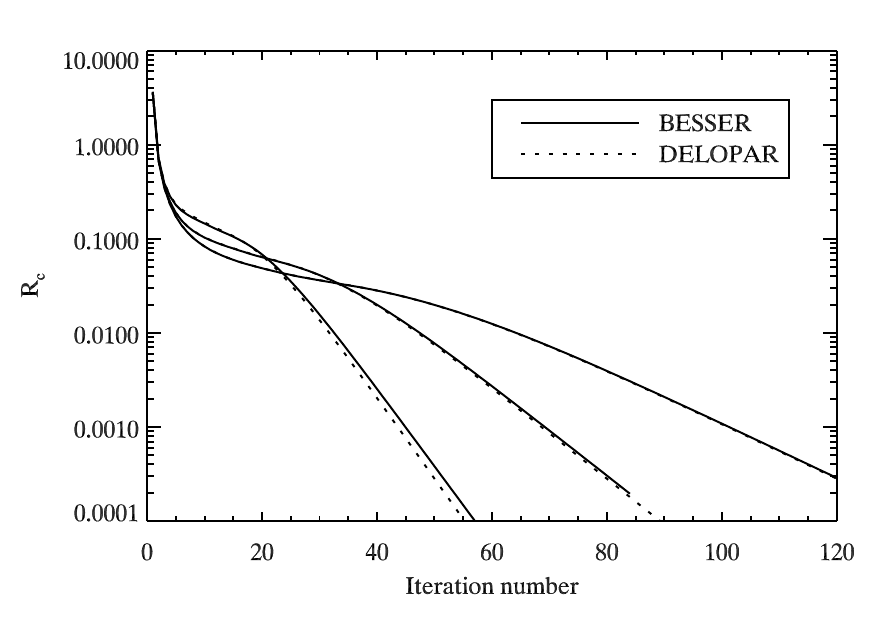}
\end{center}
\caption{
Each pair of curves shows, for a spatial grid of given resolution,  
the maximum relative change versus the iteration number using the Jacobi method with DELOPAR (dotted lines) and with BESSER (solid lines). We note that the finer the grid the slower the convergence rate, but that for any given spatial resolution both formal solvers give the same convergence rate.
}
\label{fig:compare}
\end{figure}

It is also important to note that the convergence rate of multilevel iterative methods using BESSER as formal solver is virtually identical to that achieved using DELOPAR (see an example of the Jacobi method in Fig.\,\ref{fig:compare}).

If the atmospheric model used is sufficiently smooth and no abrupt changes of the source function are present, the accuracy of BESSER and DELOPAR are virtually identical. This is not surprising because both formal solvers produce identical results in the absence of overshoots.

\subsection{Accuracy of the BESSER formal solver\label{ssec:besseraccur}}

\begin{figure}
\begin{center}
\includegraphics{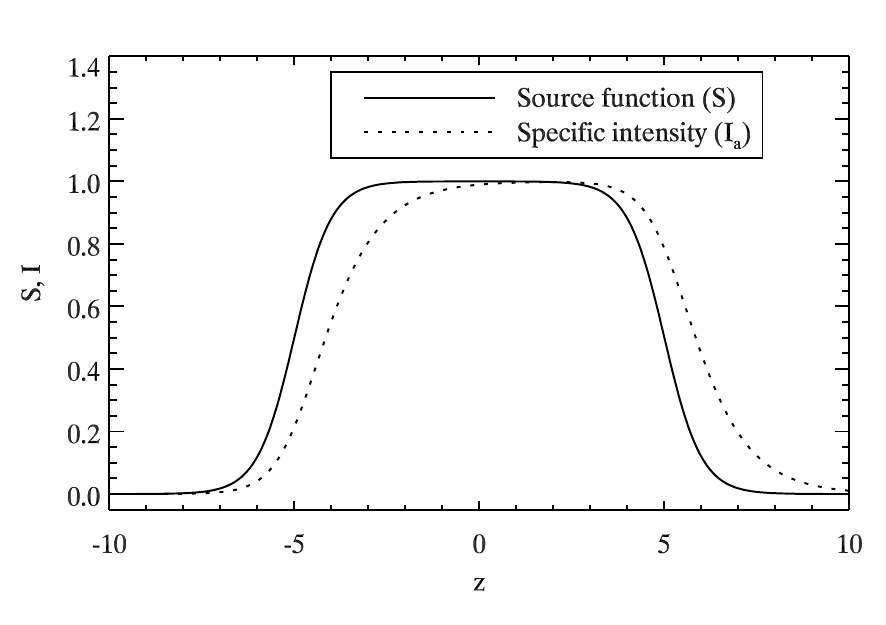}
\end{center}
\caption{
Variation along the ray direction of the source function (Eq.\,\ref{eq:ssigmas}) and of the corresponding specific intensity (Eq.\,\ref{eq:analyti}) calculated analytically.
}
\label{fig:asf}
\end{figure}

To demonstrate the accuracy of our BESSER formal solver we consider the RT problem of an arbitrary ray propagating in an infinite medium having constant opacity and a source function variation along the ray direction given by the expression (for the sake of simplicity we consider the unpolarized case)
\begin{equation}
S(z)=\sigma(2,-5,z)\sigma(-2,5,z)\,,
\label{eq:ssigmas}
\end{equation}
where the sigmoid function $\sigma$ reads
\begin{equation}
\sigma(a,d,z)=\frac{1}{1+e^{-a(z-d)}}\,
\label{eq:sigma}
\end{equation}
and $z$ is the geometrical distance along the ray, measured in units of the length scale for which $\Delta{z}=\Delta{\tau}$, with $\Delta{z}$ the grid spacing and $\Delta{\tau}$ the ensuing optical distance. As shown by the solid line in Fig.\,\ref{fig:asf}, which corresponds to $d=5$ in Eq.\,(\ref{eq:sigma}), the source function exponentially rises around $z= -d$, reaches its maximum value around $z=0$, and then exponentially decreases around $z=d$. Assuming $I_a(-\infty)=0$, the analytical solution of the radiative transfer equation for the specific intensity propagating towards positive $z$ values is (see the dotted line in Fig.\,\ref{fig:asf})
\begin{equation}
I_a(z)=
\frac{e^{15-z}}{e^{20}-1}
\left(e^{10} {\rm arctan}\,e^{z-5}-{\rm arctan}\,e^{z+5}\right)\,.
\label{eq:analyti}
\end{equation}

We have calculated numerically the specific intensity for the above-mentioned one-ray problem by solving the radiative transfer equation in several spatial grids of increasing resolution and using various formal solvers. Our aim is to compare the accuracy of our BESSER formal solver with other short-characteristics methods. To this end, we use the analytical solution given by Eq.\,(\ref{eq:analyti}) to compute the maximum true error
\begin{equation}
E(\Delta \tau)={\rm Max}\,{\Big |} \frac{I(z)-I_a(z)}{I_a(z)} {\Big |}\, ,
\end{equation}
among all the spatial points along the ray for which the solution has been obtained (i.e., $-12\,{\le}\,{z}\,{\le}\,12$).\footnote{In the numerical calculation, we use the boundary intensity $I(-12)=I_a(-12)\approx 2.77176\times 10^{-7}$.} 

\begin{figure}
\begin{center}
\includegraphics{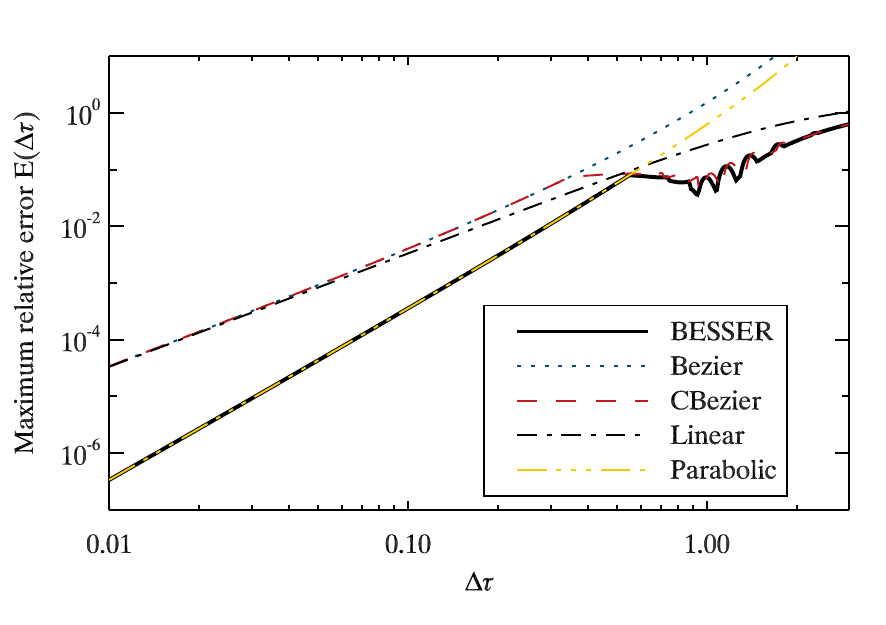}
\end{center}
\caption{
Maximum relative true error $E(\Delta\tau)$ calculated as a function of the uniform grid spacing $\Delta\tau$, using different formal solvers. Solid line: our BESSER method. Dotted line: quadratic B\'ezier with the derivative at point O calculated using the expression given by \citet{fritsch84} \citep[see also Eq. 12 of][]{auer03}. Dashed line: as in the previous method, but applying the $c_{\rm M}$ overshoot correction \citep[see Eq. 11 of][]{auer03}. Dashed-dotted line: standard SC method with linear interpolation. Three-dotted-dashed line: standard SC method with parabolic interpolation.
}
\label{fig:amethods}
\end{figure}

The formal solvers we have applied are listed in the caption of Fig.\,\ref{fig:amethods}, which gives $E(\Delta\tau)$ as a function of the $\Delta{\tau}$ of the grid spacing. Surprisingly, the worst performance is that corresponding to the B\'ezier formal solver based on the central-point derivative $y'_{\rm O}$, calculated using the weighted harmonic mean derivatives of \citet{fritsch84} and ignoring the overshoot test in the upwind interval (Bezier, dotted line). If the correction to the upwind overshoot is applied, the method performs much better, at least in the coarsest grids in Fig.\,\ref{fig:amethods} (CBezier, dashed line). In finer grids, however, the accuracy is still lower than that of BESSER and even than that provided by the standard SC method with parabolic interpolation (dashed-three-dotted line). The reason is that the estimation of the central-point derivative $y'_{\rm O}$ provided by \citet{fritsch84} \citep[see also Eq. 12 of][]{auer03} generally does not allow the second-order polynomials to be interpolated exactly.\footnote{We point out, however, that the quadratic Bezier method seems to provide reliable results in some cases of practical interest \citep{jaime13}.}

In the coarsest grids in Fig.\,\ref{fig:amethods}, the maximum true error depends not only on the grid spacing but also on the particular position of the grid points with respect to the $z=0$ position of the source function maximum. Consequently, in the region in Fig.\,\ref{fig:amethods} corresponding to the coarsest grids we observed an oscillatory behavior of the maximum true error. However, the overall variation of the error with $\Delta\tau$ remains the same independent of the particular location of the grid nodes.

\section{Parallelization using The Snake Algorithm\label{sec:snake}}

The slowest part in the numerical computations needed for solving a non-LTE 
radiative transfer problem is the formal solution of the radiative transfer equation because the number of floating point operations needed to compute the radiation field at all the spatial grid points far exceeds the number of operations needed to solve the SEEs. In particular, an accurate modeling of the spectral line polarization produced by anisotropic radiation pumping (scattering line polarization) and its modification by the Hanle effect requires the use of very fine frequency and direction quadratures that increase the computing time of the formal solution.

The formal solution for computing a single Stokes parameter at any given grid point for any given frequency and ray direction typically takes about 1\,$\mu$s on today's workstations. It is easy to estimate that one formal solution for computing the four Stokes parameters in a 3D grid with $500^3$ points, 100 radiation frequencies, and $160$ discrete directions will take about 90 days on the same computer. The full non-LTE solution requiring one hundred Jacobi iterations would take 25 years.

The use of powerful methods for multilevel radiative transfer applications, such as the non-linear multigrid method proposed by \citet{pfb97}, is necessary but not sufficient for doing multilevel radiative transfer calculations in realistic 3D atmospheric models resulting from state-of-the-art MHD simulations. To this end, we need to use massively parallel computers, which requires a suitable parallelization of the RT code. In this section, we describe a novel algorithm for performing the numerical formal solution using multiple CPU cores. As shown below, simultaneous parallelization via domain decomposition and parallelization in the frequency domain results in a very efficient radiative transfer code that shows an optimal scaling with the number of CPU cores.

\subsection{Domain decomposition\label{ssec:ddec}}

\begin{figure}
\begin{center}
\includegraphics{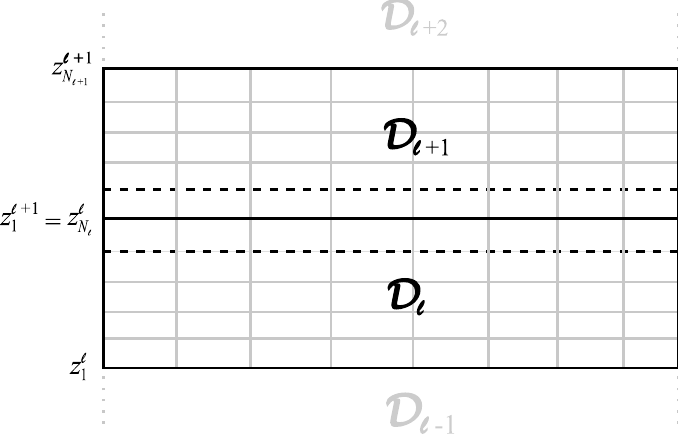}
\end{center}
\caption{
Domain decomposition in the $z$-axis, with $N_\ell$ denoting the number of discrete heights within  domain $\mathcal{D}_{\ell}$. The solid line $z^\ell_{N_\ell}=z^{\ell+1}_1$ indicates the boundary layer of the domains $\mathcal{D}_\ell$ and $\mathcal{D}_{\ell+1}$, while the dashed lines indicate the ghost layers $z_{N_\ell-1}^\ell$ and $z^{\ell+1}_2$.
}
\label{fig:domains}
\end{figure}

The computer memory needed tp store large model grids exceeds the capacity of the computing nodes of today's supercomputers by at least one order of magnitude. The capacity of computers will continue to increase in the future, but the same will happen with the scale and resolution of the MHD models. It is therefore necessary to reduce the memory demands per CPU core.  This can be achieved through the technique of domain decomposition, by means of which different parts of the model grid are treated simultaneously (in parallel), each running on a different CPU core. This task is non-trivial in radiative transfer because of the need to use a well-defined sequence of grid points. This complicates the treatment of radiative transfer in generally decomposed grids where different parts are to be solved simultaneously. A possible solution to this problem comes from the fact that the non-LTE problem needs to be solved by iteration and the radiation field at the domain boundaries can be fixed in every iteration using the radiation field calculated in the previous iteration. After a sufficient number of iterations, a self-consistent solution to the problem is eventually found. The disadvantage of this approach is that fixing the boundary radiation field of the domains reduces the information flow between the domains, which leads to a scaling of the algorithm proportional to $P^{2/3}$, with $P$ the number of CPU cores \citep{hayek08}. In Sect.~\ref{ssec:advantage} we discuss the advantages and disadvantages of our approach.

Given a Cartesian grid with $N_xN_yN_z$ discrete points, we only divide it along the $z$-axis into a consecutive sequence of $L$ domains $\mathcal{D}_\ell$, $\ell=1,\dots,L$ (see Fig.\,\ref{fig:domains}, which shows a 2D instead of a 3D grid, for simplicity). The horizontal extension of each domain $\mathcal{D}_\ell$ is always the same, and identical to that corresponding to a serial solution of the same non-LTE problem. Each of these domains is treated by one or more CPU cores in parallel with others according to the algorithm described in the following section. The boundary layer $z^\ell_{N_\ell}=z^{\ell+1}_1$ of the successive domains $\mathcal{D}_\ell$ and $\mathcal{D}_{\ell+1}$ has to be taken into account in each of the domains. Ghost layers have to be included in both domains if the formal solver of the transfer equation is of parabolic accuracy and/or the multigrid method is used. This ghost layer is needed to calculate the radiation transfer coefficients at the downwind point P (see Fig.~\ref{fig:stencil}) when point O is in the boundary layer $z^\ell_{N_\ell}=z^{\ell+1}_1$. We note that if the interpolation of the upwind and downwind radiation transfer coefficients is bilinear instead of biquadratic, only one ghost layer is needed for each of these domains. This is usually a good approximation given the spatial fineness of today's MHD models. Given that the boundary layer that is common to each pair of domains has to be treated twice, the number $N_\ell$ of $z$-points per domain is equal to $(N_z-1)/L+1$ (assuming that $L$ is such that $(N_z-1)/L$ is an integer).

As shown below, it is possible to divide the $z$-axis into a large number of intervals without any serious effect on the efficiency. In the numerical experiments discussed below, we have used values as small as $(N_z-1)/L+1=6$. For a sufficient number of computing nodes, it follows that the memory requirements per domain scales as $O(N_xN_y)$. Given the large spatial extension of the 3D stellar atmospheric models that result from today's MHD simulations \citep[e.g.,][]{jor12}, the domain decomposition strategy described above is very suitable for reducing to reasonable values the memory requirements per computing core.

\subsection{3D formal solution in the domain-decomposed models: The Snake Algorithm\label{ssec:snake}}

\begin{figure*}
\begin{center}
\includegraphics[width=15cm]{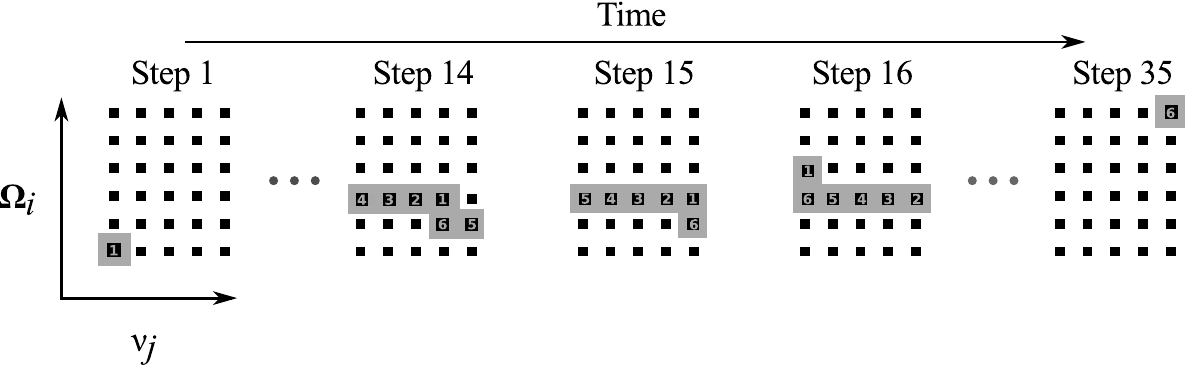}
\end{center}
\caption{
Clarification of the Snake Algorithm (SA) using an example of a formal solution with five radiation frequencies $\nu_{j=1,\dots,5}$ and six directions $\vec\Omega_{i=1,\dots,6}$, running in a domain-decomposed grid with six domains. We note that only the solution for rays $\Omega_z>0$ is shown in this figure. See the main text for details.
}
\label{fig:ppa}
\end{figure*}

In contrast to the usual 3D domain decomposition technique, it is possible to fulfill the requirement of a topologically sorted grid without the need to iterate the boundary conditions. For reasons that will become obvious below, we call it the Snake Algorithm (SA). It proceeds as follows.

The formal solution of the RT equation allows us to obtain, at each spatial grid point $(i_x,i_y,i_z)$ of domain $\mathcal{D}_\ell$, the Stokes parameters for all the discretized directions and radiation frequencies $(\vec\Omega_i,\nu_j)$. The total number of these points is $N_\Omega N_\nu$, where $N_\Omega$ denotes the number of ray directions and $N_\nu$ the number of radiation frequencies. Without loss of generality, let us consider the formal solution of the RT equation in domain $\mathcal{D}_\ell$ for the directions $\vec\Omega_i$ having $\Omega_z>0$ (i.e., for rays propagating along directions of increasing $z$, from the lower boundary of $\mathcal{D}_1$ to the upper boundary of $\mathcal{D}_L$).

For $(\vec\Omega_1,\nu_1)$, we solve the RT equation starting at the lower boundary $z^1_1$ and proceeding upwards to the domain boundary layer $z^1_{N_1}$ (see Fig.~\ref{fig:domains}). If the atmospheric model assumes periodic boundary conditions in the horizontal $(x,y)$ directions, we take them into account following the strategy of Auer et al. (1994). Since the domain is not decomposed in the horizontal directions, in each of the domains our algorithm works exactly as it does in the serial solution.

Once the radiation field for $(\vec\Omega_1,\nu_1)$ is known at the last plane $z^1_{N_1}$ of the domain, we start the process responsible for doing the formal solution in the next domain $\mathcal{D}_2$. In addition to the Stokes parameters, the $\Lambda^*_{\vec\Omega_1\nu_1}$ value has to be provided to the next domain. At this point, the process $\mathcal{D}_1$ starts solving the radiative transfer equation for $(\vec\Omega_1,\nu_2)$, beginning again at the lower boundary. After reaching the $z^1_{N_1}$ plane, the radiative data are provided to the $\mathcal{D}_2$ domain and the solution continues with $(\vec\Omega_1,\nu_3)$. These steps are repeated until the radiation transfer equation is solved for all the discrete frequencies. Then, it continues in an analogous way with $(\vec\Omega_2,\nu_1)$ and for all the directions with $\Omega_z>0$.

The solution in domain $\mathcal{D}_2$ proceeds in an exactly analogous way. After receiving the radiation data $(\vec\Omega_1,\nu_1)$ from domain $\mathcal{D}_1$, the RT equation is solved in planes $z^2_2$, $z^2_3$, etc., up to the layer $z^2_{N_2}$, from which the resulting radiation field and $\Lambda^*_{\vec\Omega_1\nu_1}(x,y,z^2_{N_2})$ are propagated to the grid points $(x,y,z^3_1)$ of domain $\mathcal{D}_3$. At a given time, each domain solves the RT equation for different $(\vec\Omega_i,\nu_j)$, such that the difference between two successive processes (domains) is just one step in the discrete space of directions and frequencies. The outgoing radiation from one domain becomes the incoming radiation for the following domain. The resulting snake of length $L$ clambers half of the parameter space of directions $\Omega_z>0$ and, after this is finished, it proceeds back in an analogous way by solving the radiative transfer problem for all $\Omega_z<0$ directions.

Figure~\ref{fig:ppa} visualizes the whole process using as an example a formal solution with five radiation frequencies $\nu_{j=1,\dots,5}$ and six directions $\vec\Omega_{i=1,\dots,6}$, running in a domain-decomposed grid with six domains, each of which is indicated by a numbered rectangle. Each step of the solution in every domain corresponds to a formal solution of the RT equation for one direction and one frequency. In the next step, the snake of processes moves by one $(\vec\Omega_i,\nu_j)$ point and the radiation field data are passed between the successive domains. In this example, every single process solves the RT problem in the dedicated domain which contains $N/6$ grid nodes. These processes parse the discrete space of directions and frequencies in the well-defined order indicated in the figure, until the whole direction-frequency space is passed through by all the processes. At the beginning and at the end of the formal solution, some of the processes are inactive, waiting for other processes to finish their work. The total number of time steps of the formal solution is 35 for the 30 direction-frequency points, which implies a speedup factor $30\times 6/35\approx 5.1$ with respect to the serial solution. This can be easily verified by using Eq.~(\ref{eq:accelz}) with $N_\Omega=12$ (see below).

If the domains have the same or similar $N_\ell$ values (which is easy to achieve in practice), each process (i.e., each domain) solves only the fraction $1/L$ of the whole radiation transfer problem. This leads, in principle, to an almost linear scaling with the number of spatial domains.

For practical reasons (i.e., optimization in the treatment of line absorption profiles, because it is not practical to store them for every grid point and ray direction), the line absorption profiles are obtained by interpolation, using a pre-calculated database created at the beginning of the non-LTE solution. This implies some significant reduction of memory and computing time. After calculating a line profile from the database (where the profiles are normalized to unit area), one may need to renormalize it using the chosen frequency quadrature (e.g., in the presence of  Doppler shift gradients caused by macroscopic plasma motions). This only needs to be done once per direction if the loop over frequencies is the inner loop because the normalization factor only depends on direction. In summary, it is more convenient if the directions are in the outer loop and the frequencies in the inner loop of the algorithm, so that our  snake parallelization strategy proceeds row by row as indicated in Fig.~\ref{fig:ppa}, instead of column by column.

Concerning the implementation of the algorithm, it is important to note that it is crucial to use non-blocking routines to propagate the radiation data between the successive domains. In other words, the RT calculation in domain $\mathcal{D}_\ell$ proceeds by solving the next $(\vec\Omega_i,\nu_j)$ point immediately after the lower-boundary radiation data from domain $\mathcal{D}_{\ell-1}$ arrives. It does not have to wait for $\mathcal{D}_{\ell+1}$ to retrieve the $(\vec\Omega_i,\nu_{j-1})$ data. Consequently, the snake in Fig.~\ref{fig:ppa} can temporarily become split, with two successive processes $\ell$ and $\ell+1$ processing non-subsequent points in the discrete $\vec\Omega\times\nu$ space (see Fig.~\ref{fig:ppa}). If this does not happen, the computing performance can decrease significantly because a significant amount of time is spent waiting for the synchronization of the whole grid.

\subsubsection{Scaling of the algorithm}

\begin{figure}
\begin{center}
\includegraphics{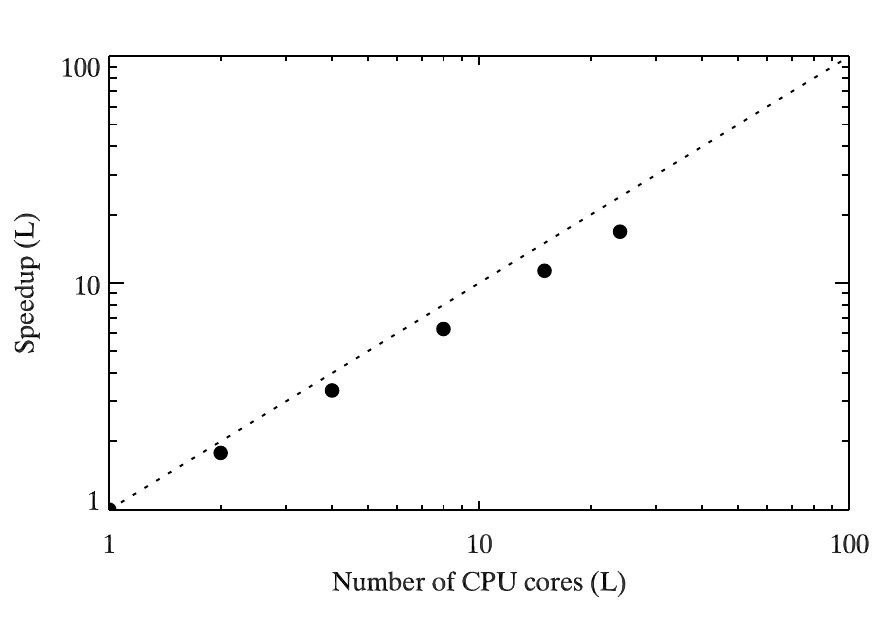}
\end{center}
\caption{
Speedup $S(L)$ of the solution of the RT equation due to domain decomposition with the Snake Algorithm. The number of CPU cores on the horizontal axis is equal to the number $L$ of spatial domains. The diagonal dotted line indicates the theoretical curve of linear scaling. The scaling of the algorithm is almost a linear function of the number of domains. The small departure from linearity is mainly due to the cost of the inter-process communication.
}
\label{fig:bench-z}
\end{figure}

In the serial solution, the computing time needed for the formal solution of the RT equation is proportional to the total number of spatial grid points $\mathcal {N}=N_xN_yN_z$, the number of directions $N_\Omega$, and the number of radiation frequencies $N_\nu$. We shall denote this computing time by
\begin{equation}
T_1=\alpha {\mathcal {N}} N_\Omega N_\nu\,,  \label{eq:serialtime}
\end{equation}
with $\alpha$ a constant of proportionality. In the domain-decomposed parallel solution (still assuming that the first half of the integrations is performed along the directions having $\Omega_z>0$), the duration of the full formal solution in the whole grid is equal to the time interval between the first ($\vec\Omega_1,\nu_1$) and the last ($\vec\Omega_{N_\Omega},\nu_{N_\nu}$) ray integrations in the domain $\mathcal{D}_1$.

Given that the number of grid nodes in domain $\mathcal{D}_1$ is equal to ${\mathcal {N}}/L$, the time spent solving the first half of the rays in domain $\mathcal{D}_1$ is
\begin{equation}
t_a=\alpha  \frac {\mathcal {N}}L \frac{N_\Omega}{2} N_\nu\,.
\end{equation}
The process responsible for domain $\mathcal{D}_1$ is waiting for the last upward ray $(\vec\Omega_{N_\Omega/2},\nu_{N_\nu})$ to propagate through $L-1$ domains to the upper grid boundary $z_{N_z}$. Its duration equals
\begin{equation}
t_b=\alpha(L-1) \frac{{\mathcal {N}}}{L}\,.
\end{equation}
The same time $t_b$ it taken by the first downward ray $(\vec\Omega_{N_\Omega/2+1},\nu_1)$ to propagate from $\mathcal{D}_L$ to the upper boundary of $\mathcal{D}_1$. The computing time needed for the solution of the $N_\Omega/2$ downward rays in the $\mathcal{D}_1$ domain is, again, equal to $t_a$.

The duration of the full formal solution in the $L$-decomposed grid is, therefore, equal to $T_L=2(t_a+t_b)$. It follows from the equations above that
\begin{equation}
T_L = \alpha\frac {\mathcal {N}}L\left[ N_\Omega N_\nu+2(L-1) \right]\,.      \label{eq:parLtime}
\end{equation}
We define the speedup of the parallel solution with respect to the serial solution as
\begin{equation}
S(L)=\frac{T_1}{T_L}\,,
\end{equation}
which, using Eqs. (\ref{eq:serialtime}) and (\ref{eq:parLtime}), is equal to
\begin{equation}
S(L)=L\frac{1}{1+\lambda}\,,
\label{eq:accelz}
\end{equation}
where
\begin{equation}
\lambda = \frac{2(L-1)}{N_\Omega N_\nu}\,.
\label{eq:lambdasnake}
\end{equation}

If $\lambda\ll 1$, then the speedup in the formal solution is practically linear with the number of domains $L$. This is equivalent to saying that $N_\Omega N_\nu\gg L$, i.e., to a situation in which the number of direction-frequency points is much larger than the number of domains. Given that in the transfer of polarized radiation the typical orders of magnitude of the relevant quantities are $N_\Omega\sim 10^2$, $N_\nu\sim 10^3$, and $L\sim 10^2$, we obtain $\lambda\sim 10^{-3}$. It is easy to see from Eq.~(\ref{eq:accelz}) that SA always accelerates the solution if $N_{\vec\Omega}>2$ and $L>1$.

\subsection{Parallelization in the radiation frequencies\label{ssec:freqpar}}

\begin{figure}
\begin{center}
\includegraphics{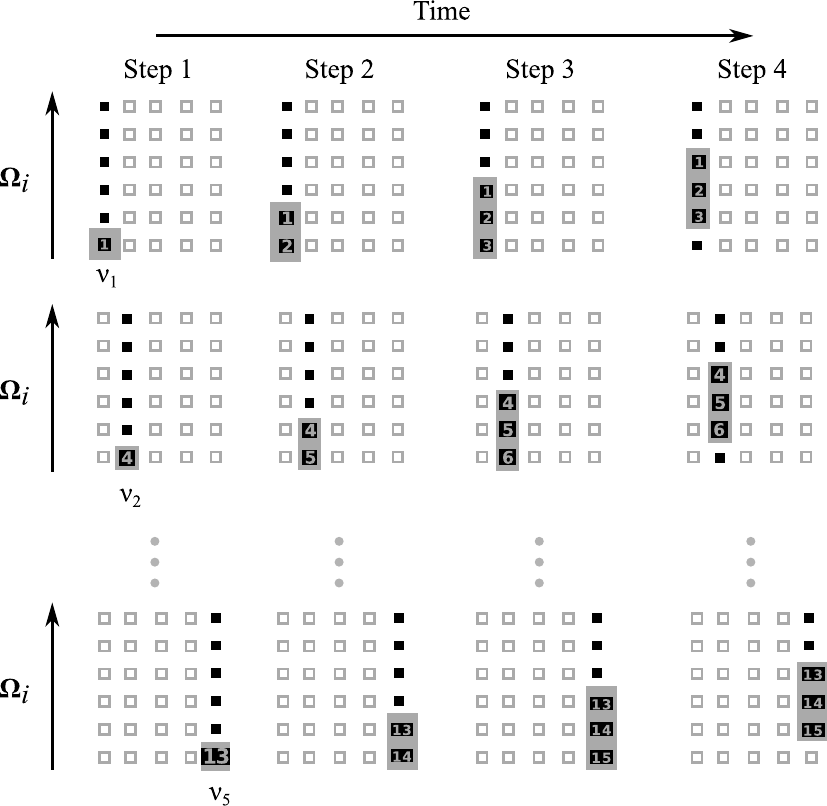}
\end{center}
\caption{The Snake Algorithm applied to the problem in Fig.~\ref{fig:ppa} with $L=3$ and $M=5$. Here, every process has a single dedicated radiation frequency, i.e., $N_m=1$. Given the small number of directions and frequencies in this illustrative example, we have $\lambda=2/3$ and the speedup $S(LM)=9$. See the text for details.
}
\label{fig:ppa2}
\end{figure}

The radiation frequencies $\nu_{i=1,\dots,N_\nu}$ can be grouped into $M$ intervals, each containing the $N_m=N_\nu/M$ discrete frequencies.\footnote{At least in the favorable situation in which $N_m$ is an integer number. In general, it is convenient that the individual frequency intervals have similar lengths.} The Snake Algorithm can be applied in parallel to each of these frequency blocks. The only difference with respect to the algorithm described in Sect.~\ref{ssec:snake} is that the solution in the spatial domains $\mathcal{D}_\ell$ is only performed in a sub-space of $(\vec\Omega_i,\nu_j)$ in which $j=j^m_1,\dots,j^m_{N_m}$. Since this can be done in parallel for all the $M$ blocks, a significant reduction of the solution time can be achieved (see Fig.~\ref{fig:ppa2}).

The domain and frequency decomposition parallelization strategies described above are performed independently of each other, in the sense that there is no need of communication between the processes treating different frequency intervals $m$ during the formal solution. The radiation field tensors $J^K_Q$ and the $\Lambda^*_{\Omega\nu}$ operator needed for the solution of the statistical equilibrium equations are only partially integrated over the line absorption profiles during the formal solution and, at the end of the whole formal solution process, are summed over the frequency intervals and synchronized among them. The time cost of this operation is negligible with respect to the time demands of the formal solution. Thanks to this orthogonality of the two independent parallelizations, it is possible to achieve a multiplicative effect of both speedup factors.

\subsubsection{Scaling of the algorithm}

\begin{figure}
\begin{center}
\includegraphics{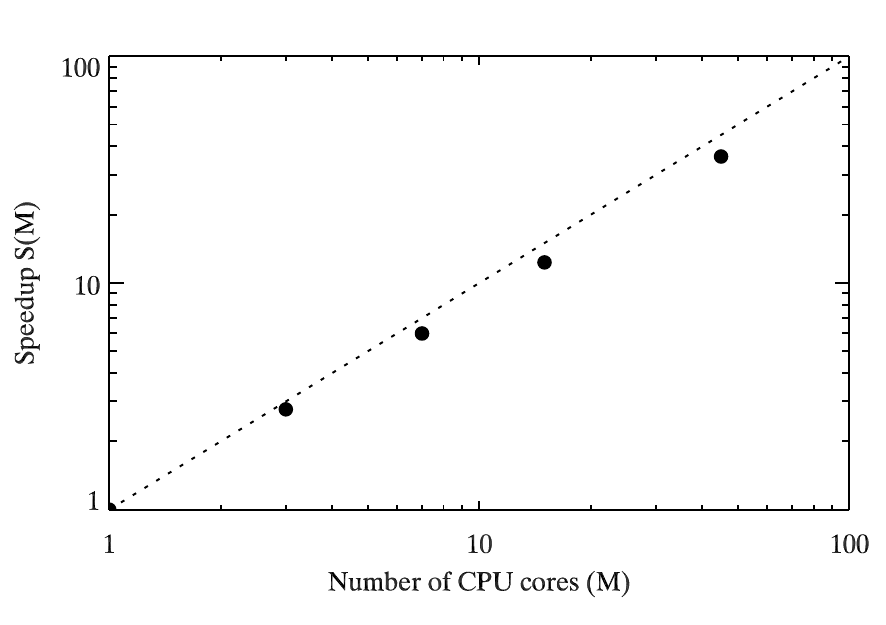}
\end{center}
\caption{
Speedup $S(M)$ of the formal solution of the RT equation due to parallelization in radiation frequencies in a single spatial domain ($L=1$).
}
\label{fig:bench-f}
\end{figure}

A reduction in the number of frequencies in every domain by a factor of $1/M$ gives a new solution time $T_{ML}$ which is obtained after the replacement $N_\nu\to N_\nu/M$ in Eq.~(\ref{eq:parLtime}). The speedup then follows as
\begin{equation}
S(ML) = M L\frac{1}{1+\lambda}\,,
\label{eq:tml}
\end{equation}
where we now have
\begin{equation}
\lambda=\frac{2M(L-1)}{N_\Omega N_\nu}\,.
\label{eq:lambdaLM}
\end{equation}
When $\lambda \ll 1$ the scaling of the algorithm is virtually linear with the total number of CPU cores, $P=ML$. As an example, we assume a large 3D model atmosphere and a supercomputer with $10^4$ CPUs which allows for parallelization with $L=100$ and $M=100$. Assuming $N_\Omega=200$ and $N_\nu = 1\,000$, we have $\lambda\approx 0.1$ and a speedup $S(ML)\approx 0.9\times 10^4$.

If $ML\gg 1$, then Eq.~(\ref{eq:tml}) can be approximated by the expression
\begin{equation}
S(ML) \approx \left[\frac 1{ML} + \frac 2{N_\Omega N_\nu}\right]^{-1}\,,
\end{equation}
which immediately shows how the speedup factor depends on both the fineness of the quadratures and the number of available CPU cores. It follows that it is only possible to decrease the computing time by an additional factor of $1/2$ when using more cores than
\begin{equation}
(ML)_{\rm optimal} = \frac {N_\Omega N_\nu}{2}\,.
\end{equation}
Given that typically $N_\Omega N_\nu\sim 10^{5}$, the power of today's supercomputing facilities can be effectively exploited. Not surprisingly, $(ML)_{\rm optimal}$ corresponds to a situation in which each frequency is treated by a unique frequency band and the number of spatial domains is equal to one half of the number of rays (see Eq.~(\ref{eq:lambdasnake}) for $N_\nu=1$).

\begin{figure}
\begin{center}
\includegraphics{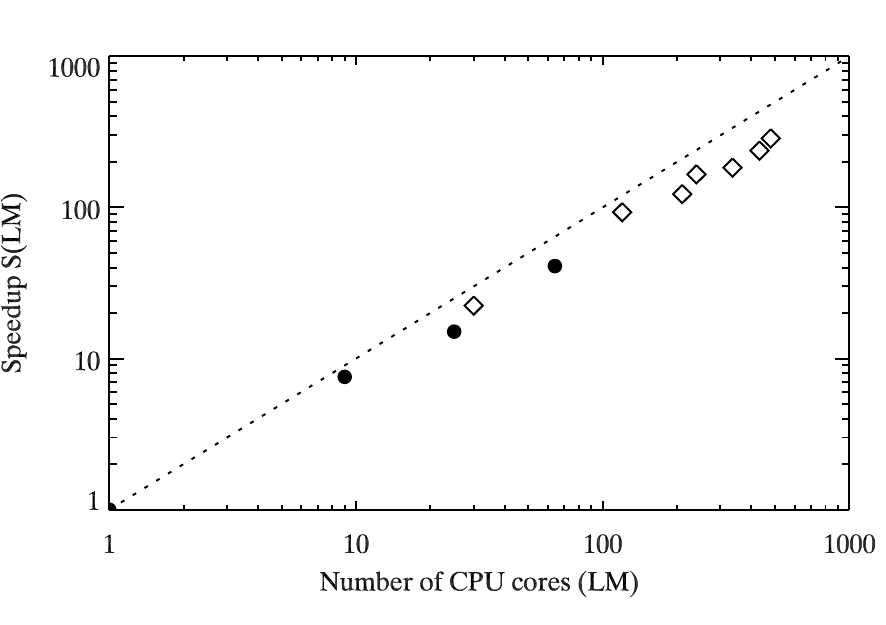}
\end{center}
\caption{
Speedup $S(LM)$ of the formal solution of the RT equation due to simultaneous parallelization in radiation frequencies and domain decomposition. Filled circles: data calculated for various values of $L$ and $M$ using the OCAS cluster. Diamonds: data calculated using the LaPalma supercomputer. We note that because of non-optimal spatial and frequency decompositions (i.e., not all the CPU cores treat exactly the same number of grid points and/or frequencies), small oscillations around the nearly linear trend appear. The speedup in both data sets is normalized to the serial time $T_1$ corresponding to each of the used computers.
}
\label{fig:bench-pf}
\end{figure}

\subsection{Advantages and disadvantages of the Snake Algorithm\label{ssec:advantage}}

The advantages and disadvantages of SA can be summarized as follows.\\

Disadvantages:
\begin{enumerate}
\item It is only possible to decompose the grid so that $L<N_z$.
\item If $N_xN_y$ becomes too large, the memory requirements can be difficult to fulfill in very large atmospheric models. In the particular case of the five-level Ca\,{\sc ii} model atom considered in this paper, a conservative memory limit of 2\,GB per core would be reached for models with $N_xN_y\approx 1500^2$.
\item If the number of ray directions and frequencies is low, i.e., $\lambda\gtrsim 1$ in Eq.~(\ref{eq:lambdaLM}), the scaling properties of SA are sub-optimal. However, this situation is unlikely in realistic models.
\item It is not trivial to implement the Gauss-Seidel iterative method of Trujillo Bueno \& Fabiani Bendicho (1995) for the smoothing part of the non-linear multigrid iteration (see Sect.\,\ref{sec:mg}).
\item There is always an overhead of communication and synchronization between the domains. This overhead can become as large as 20\,\% of the whole formal solution computing time for cases with $N_z/L$ approaching unity,  but the overhead is usually about 10\,\% with the hardware we have been using (see Appendix~\ref{app:bench}). The relative importance of the cost of the inter-process communication increases with the number of domains, leading to a minor departure from linear scaling in the parallel solution.
\end{enumerate}

Advantages:
\begin{enumerate}
\item The accuracy of the solution is identical to the serial solution. The following two points follow directly from this fact.
\item The total number of iterations needed to solve the non-LTE problem 
is equal to the serial solution and the convergence rate is not deteriorated as in domain-decomposition methods with iterative boundary conditions.
\item There are no discontinuities of the errors at the boundaries of domains due to iterative boundary conditions. The error remains smooth and the application of the non-linear multigrid method is not affected by the presence of such numerical problems.
\item It is easy to use a partially or fully converged solution (stored in a disk file) as an initialization of a different non-LTE computation without having to worry about the radiation field at the domain boundaries.\footnote{We note that to store the full radiation field would be practically impossible given the number of radiation field quantities in polarization transfer problems.} For instance, this is useful for obtaining the solution in a given 3D thermal model, but for different magnetic field choices.
\item The scaling properties of the algorithm are almost linear with the number of CPU cores $P$, until $P$ becomes comparable to  $N_\Omega N_\nu$, i.e., for all cases of practical interest.
\item Thanks to the multiplicative effect of the two independent parallelization strategies (i.e., domain decomposition and parallelization in radiation frequencies), large-scale supercomputing facilities can be used with a significant improvement in the solution time.
\item The algorithm is relatively easy to implement in practice. Existing serial RT codes can be generalized using the SA without a very serious programming effort.
\end{enumerate}

\section{The non-linear multigrid method\label{sec:mg}}

As mentioned in Sect.\,\ref{sec:formul} we need a fast iterative method capable of finding rapidly the density matrix elements $\vec\rho_l$ such that Eq.~(\ref{eq:nonlte}) is satisfied when the radiative rates, which appear in the block-diagonal matrix $\bm{\mathcal{L}}_{l}$ of Eq.~(\ref{eq:nonlte}), are calculated through the solution of the Stokes-vector transfer equation when using at each spatial point the emission vector and the propagation matrix corresponding to these $\vec\rho_l$ values. A suitable method is the Jacobian iterative scheme described in \citet{rms03} and in Appendix\,A of \citet{stepan11}. As with other operator splitting methods, with this Jacobi-based iterative method the convergence rate is the slower the finer the spatial grid, so that the computing time needed to obtain the self-consistent solution scales in general as ${{{\mathcal {N}}}_l}^2$ (where ${{{\mathcal {N}}}_l}$ is the total number of spatial grid points). In order to solve complicated 3D problems in very fine grids it is convenient to apply an iterative method whose convergence rate is insensitive to the grid size, so that the computing time needed to obtain the self-consistent solution scales as ${{{\mathcal {N}}}_l}$. This method is called the non-linear multigrid method \citep[e.g.,][]{hack85} whose application to multilevel radiative transfer without polarization has been described in great detail by \citet{pfb97}.\footnote{For the linear multigrid method, valid only for the two-level atom case, see \citet{steiner91}.} Here we provide only a brief overview of the non-linear multigrid (MG) method with emphasis on the details related to our generalization to the polarized case and the parallelization strategy via domain decomposition.

\subsection{Brief overview of the non-linear MG method}

As mentioned above, the aim is to find the $\vec\rho_l$ vector of density matrix values, defined in the desired fine grid of resolution level $l$, such that Eq.~(\ref{eq:nonlte}) is satisfied as explained above. We assume that we have a fine grid estimate $\vec\rho_l^{\rm old}$ such that the residual
\begin{equation}
{\vec r}_l={\vec f}_l-{\bm{\mathcal{L}}^{\rm old}_l}\vec\rho_l^{\rm old}{\,},
\label{eq:resid}
\end{equation}
is smooth.\footnote{We note that by writing $\bm{\mathcal{L}}^{\rm old}_l$ we want to point out that the radiative rates that appear in the block-diagonal matrix $\bm{\mathcal{L}}_l$ of Eq. (2) are calculated using the current estimate $\vec\rho_l^{\rm old}$.} We would like to obtain a fine-grid correction $\Delta\vec\rho_l$ such that the new estimate $\vec\rho_l^{\rm new}=\vec\rho_l^{\rm old}+\Delta\vec\rho_l$ satisfies Eq.~(\ref{eq:nonlte}): 
\begin{equation}
{\bm{\mathcal{L}}_l}\,\vec\rho_l^{\rm new} ={\vec f}_l.
\label{eq:mgcorr}
\end{equation}
Given that the problem is non-linear (i.e., the operator $\bm{\mathcal{L}}_l$ depends on $\vec\rho_l^{\rm new}$) we need to make an approximation to $\bm{\mathcal{L}}_l$ in order to obtain a better estimate $\vec\rho_l^{\rm new}$. 

With the Jacobian method we simplify the operator $\bm{\mathcal{L}}_l$ in the same grid of resolution level $l$; we do this by choosing the diagonal of the $\vec\Lambda$ operator. With the non-linear MG method we approximate the operator $\bm{\mathcal{L}}_l$ by forming a suitable approximation in a coarser grid of level $l-1$; that is, we coarsify rather than simplify. In order to understand how we coarsify we note first that Eqs.~(\ref{eq:resid}) and (\ref{eq:mgcorr}) imply that
\begin{equation}
{\bm{\mathcal{L}}_l}\vec\rho_l^{\rm new}\,-\,{\bm{\mathcal{L}}^{\rm old}_l}\vec\rho_l^{\rm old}={\vec r}_l\,.
\label{eq:mgcsf}
\end{equation}
The residual ${\vec r}_l$ was assumed to be smooth, so we can map the left-hand side of Eq.~(\ref{eq:mgcsf}) to a coarser grid of level $l-1$ to obtain the coarse-grid equation \citep[cf.][]{pfb97}:
\begin{equation}
{\bm{\mathcal{L}}_{l-1}}{\overline{\vec\rho}_{l-1}}={\bm{\mathcal{L}}^{\rm old}_{l-1}}{\bm{\mathcal R}}(\vec\rho_l^{\rm old})
+{\bm{\mathcal R}}({\vec r}_l)\,,
\label{eq:mgcsfy}
\end{equation}
where the linear operator $\bm{\mathcal R}$ is a fine-to-coarse or restriction operator whose application to $\vec\rho_l^{\rm old}$ and to ${\vec r}_l$ allow us to obtain directly the {\em rhs} terms. Therefore, the system of Eq.~ (\ref{eq:mgcsfy}) is formally identical to the original system of Eq.~(\ref{eq:nonlte}), but formulated in a grid of level $l-1$. After solving it with the Jacobian method explained in Appendix\,A of \citet{stepan11} we obtain the coarse-grid correction (CGC)
\begin{equation}
\vec\rho_l^{\rm new}=\vec\rho_l^{\rm old}\,+\,\bm{\mathcal P}[{\overline{\vec\rho}_{l-1}}\,-\,\bm{\mathcal R}(\vec\rho_l^{\rm old})],
\end{equation}
where $\bm{\mathcal P}$ is a coarse-to-fine or prolongation operator. As shown in Sect.\,2.3 of \citet{pfb97} the CGC is very efficient in reducing the amplitude of the low-frequency components of the error, but not the high-frequency components with wavelengths smaller than or similar to twice the distance between the coarse grid points. To achieve a convergent two-grid iterative scheme it is crucial to apply a number of iterations in the fine grid capable of removing the high frequency components of the error (smoothing iterations). \citet{pfb97} showed that their Multilevel Gauss-Seidel (MUGA) iterative method (see also Trujillo Bueno \& Fabiani Bendicho 1995) has excellent smoothing capabilities in a wide range of spatial grids, but they also showed that a multilevel iterative scheme based on the Jacobi method has similar smoothing capabilities in fine grids (e.g., with grid spacings smaller than about 50\,km in the case of the solar Ca\,{\sc ii} problem outlined in Appendix\,\ref{app:atom}). For this reason, but mainly because the implementation of a Jacobian method for massively parallel computing is relatively straightforward, the smoothing iterations in our 3D code PORTA are done applying the Jacobian method explained in Appendix\,A of \citet{stepan11}.

\begin{figure}
\begin{center}
\includegraphics[width=0.8\columnwidth]{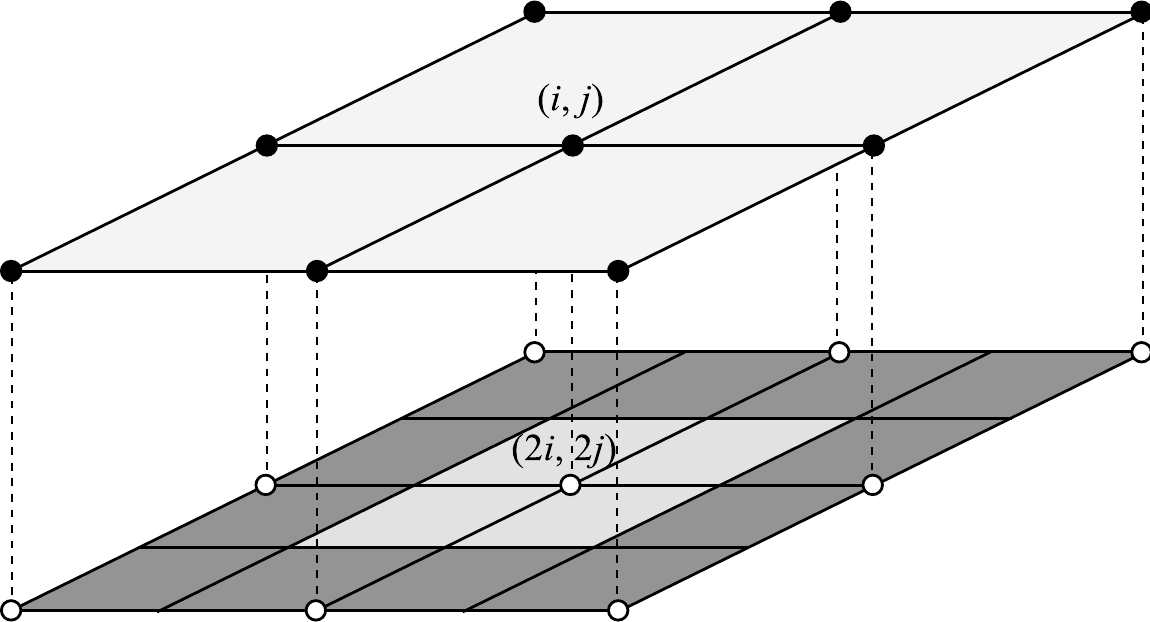}
\end{center}
\caption{
Restriction of fine-grid points (see the bottom plane) to coarse-grid ones (see the top plane).
Restriction to the coarse-grid node $(i,j)$ is performed using the data of nine fine-grid nodes making up a $3\times 3$ stencil, indicated by the brighter area of the bottom plane, with the central $(2i,2j)$ grid node. See the text for details.
}
\label{fig:mg}
\end{figure}

The previous steps correspond to a two-grid iteration: smoothing in the desired fine grid, restriction to the selected coarse grid, coarse-grid correction, prolongation to the fine grid, and smoothing in the fine grid. Three-grid and other multigrid methods can be obtained as a direct generalization of the two-grid method. The most sophisticated multigrid method is the nested multigrid method explained in Sect.\,5 of \citet{pfb97}. All these options are available in our PORTA code, where the grid of resolution level $l-1$ is derived from the grid of resolution level $l$ by removing half of the grid points corresponding to each axis, so that every coarse-grid node $(i_x,i_y,i_z)_{l-1}$ coincides with some fine-grid node $(2i_x,2i_y,2i_z)_l$ (see Fig.~\ref{fig:mg}).

\subsection{Prolongation and restriction\label{ssec:restprol}}

\begin{figure}
\begin{center}
\includegraphics[width=\columnwidth]{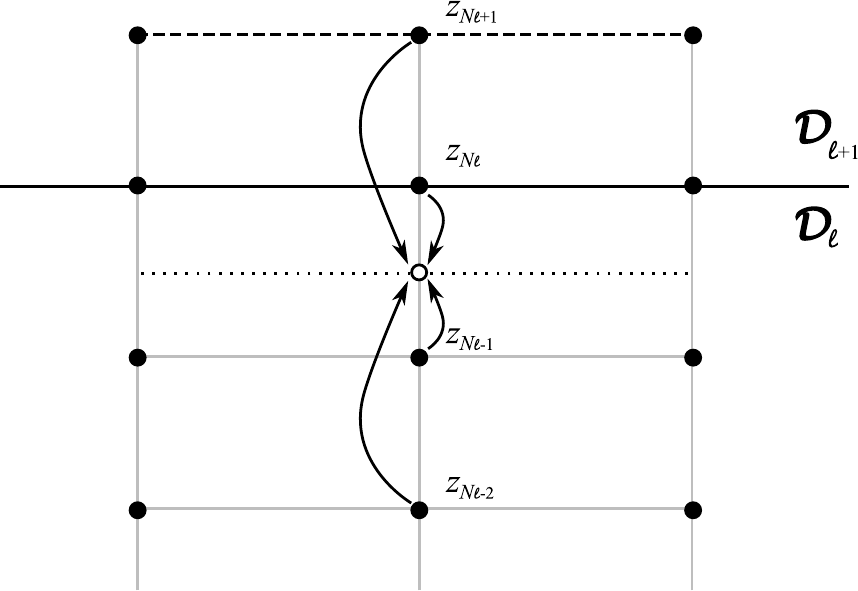}
\end{center}
\caption{Illustration of the prolongation of the coarse-grid correction 
to the fine grid using data from the ghost layer. Solid black line: the boundary layer between the domains $\mathcal{D}_\ell$ and $\mathcal{D}_{\ell+1}$. Dashed line: the ghost layer of the domain $\mathcal{D}_\ell$. Dotted line and the empty circle: the fine-grid layer near to the domain boundary and a fine-grid node. Gray lines: lines of the coarse grid. The arrows show how the data are interpolated to the fine-grid node using the cubic-centered interpolation of the coarse-grid nodes data.
}
\label{fig:mg-boundary}
\end{figure}

In PORTA, two different prolongation operators are currently implemented: tri-linear and tri-cubic-centered interpolation.

The tri-linear interpolation uses the information from eight coarse-grid nodes surrounding the fine-grid node. If the fine-grid node is located in a plane or on a grid line formed by coarse-grid nodes, bilinear or linear interpolation is used, respectively. Finally, if the fine-grid node coincides with a coarse-grid node, direct injection of the correction is applied.

Similarly, in the cubic-centered interpolation, fine-grid node information is obtained by interpolation of data from $4^3$ nearby coarse-grid nodes. Bicubic or cubic interpolations, respectively, are used for fine-grid nodes located at the plane or at a line connecting coarse-grid nodes. For tri-cubic interpolation near the boundary separating two domains of the domain-decomposed grid, data from ghost layers have to be used (see Fig.~\ref{fig:mg-boundary}). According to the rules described above, interpolation of the fine-grid data at the very boundary layer does not require any data from the ghost nodes.

Near the grid boundaries, that is, near the real boundary of the model where sufficient data are missing because of non-existent outer grid nodes, we use the data from the inner grid nodes to perform the cubic interpolation (i.e., the point of interpolation is located in one of the boundary intervals of the four-point interpolant). For this reason, the minimum number of coarse-grid nodes per axis in a domain is limited to four if the MG method with cubic interpolation is used.

The MG method with a cubic prolongation operator generally performed better in our convergence tests. On the other hand, it is less numerically stable because it may produce unphysical extrema of the density matrix components in the fine grid. This can happen in models containing abrupt changes of the physical quantities, such as in atmospheric models resulting from today's MHD simulations.
If such a situation is encountered, it is safer to use a prolongation operator with linear accuracy or to implement a monotonity-preserving B\'ezier interpolation (however, we have found the liner interpolation only slightly worse than the cubic one, and so this effort does not seem to be justified).

For the restriction operator, we use the adjoint of the tri-linear prolongation operator \citep{nr07} using the information from $3^3$ fine-grid nodes surrounding the coarse-grid node (see an analogous 2D example in Fig.~\ref{fig:mg}). These fine-grid nodes are always available even in the domain-decomposed grids because of the presence of ghost layers.
In the case of a coarse-grid node located at a non-periodic boundary of the grid, the restriction only takes into account $2\cdot 3^2$ or $2^3$ nearby fine-grid nodes depending on the particular grid topology.

We note that the ghost layer is the nearest layer to the domain boundary in any particular grid $G_l$ of resolution level $l$. Therefore, the ghost layers in the grids $G_l$ and $G_{l-1}$ correspond to layers having different coordinates $z$. On the other hand, the boundary layers between two successive domains $\mathcal{D}_\ell$ and $\mathcal{D}_{\ell+1}$ must always coincide in both discretizations $G_l$ and $G_{l-1}$.

Given that the formal solution of the RT equation and the solution of the statistical equilibrium equations in every domain $\mathcal{D}_\ell$ is only performed in the real internal grid nodes, the data in the ghost layers need to be synchronized with the neighboring domains after every solution of such equations, calculation of the residuum \citep[see Eq.~10 of][]{pfb97}, or restriction or prolongation operation.

As in the case of iterations based on the Jacobi method, the multigrid solution obtained with our parallelization strategy has exactly the same accuracy as the corresponding serial solution. The existence of domain boundaries does not destroy the high-frequency smoothing that is so crucial for the convergence of the multigrid iteration.

\subsection{Convergence of the multigrid method}

\begin{figure}
\begin{center}
\includegraphics[width=0.9\columnwidth]{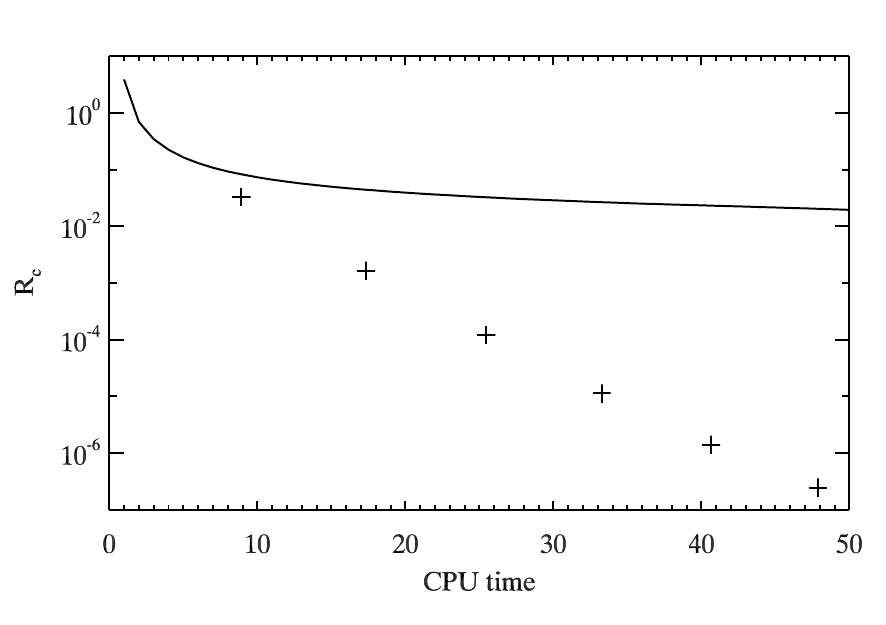}
\end{center}
\caption{
Comparison of Jacobian and MG convergence quantified by the maximum relative change of populations between iterations $R_{\rm c}$. Solid line: Jacobi iteration. Crosses: standard multigrid method with the V-cycles with two pre-smoothing and four post-smoothing iterations. The computational time on the horizontal axis is measured in units of one Jacobi iteration in the finest grid.
}
\label{fig:mg-conv}
\end{figure}

Since the smoothing capabilities of the Jacobi iteration are inferior to the Gauss-Seidel iteration, one usually needs to apply more Jacobi smoothing sweeps in order to reduce the high-frequency components of the error. We have found by numerical experimentation that two pre-smoothing and four post-smoothing iterations in the V-cycle MG method lead to optimal convergence in our atomic and atmospheric models. Given that with the MG method the convergence error $C_{\rm e}$ is lower than the maximum relative change \citep[see Eq.~19 of][]{pfb97}, it is usually sufficient to apply only two or three V-cycles to reach the convergence.

We have applied the multigrid algorithm to the five-level Ca\,{\sc ii} problem described in Appendix~\ref{app:atom} using a sequence of three grids: the finest grid $G_3$ with $100\times 100\times 121$ points, $G_2$ with $50\times 50\times 61$ points, and the coarsest grid $G_1$ with $25\times 25\times 31$ points using our domain decomposition ($L=5$) and our frequency parallelization ($M=9$).

In Fig.~\ref{fig:mg-conv}, we show a comparison of the convergence properties of the Jacobi and standard multigrid iteration. In comparison with the multigrid method based on Gauss-Seidel smoothing iterations, our multigrid method based on Jacobi smoothing iterations is slower by about a factor of three \citep[cf. Fig.~15 of][]{pfb97}. Asymptotically, both methods are proportional to each other and the deficiency of the Jacobi smoothing sweeps can be compensated for by increasing the number of computing nodes.

Finally, we want to remark that the multigrid method is a highly convergent iterative scheme, especially suitable for obtaining the self-consistent solution in very fine grids. However, care must be taken if the coarse grids used become too coarse to represent properly the physical reality of the transfer problem under consideration (i.e., if the number of grid points per decade of the optical depth becomes close to or smaller than unity). In this case the multigrid method may fail to converge.

\section{Concluding comments\label{sec:con}}

The computer program we have described in this paper, PORTA, is a powerful multilevel radiative transfer code for the simulation of the intensity and polarization of the spectral line radiation produced by scattering processes and the Hanle and Zeeman effects in 3D models of stellar atmospheres. It is based on the non-linear multigrid method proposed by Fabiani Bendicho et al. (1997) and on a new 3D formal solver of the Stokes-vector transfer equation which uses B\'ezier monotonic interpolation along short-characteristics in each radiation beam. The PORTA program can easily do Jacobi iterations in the desired spatial grid, instead of solving the non-LTE problem under consideration through the non-linear multigrid method and has two general options: one for spectroscopy (for those interested only in the familiar intensity spectrum) and another one for spectropolarimetry (for those interested in the full diagnostic content of the spectral line radiation). The benchmarks we have carried out up to now and the first applications to spatially complex models of the extended solar atmosphere \citep{stepan12} indicate that PORTA is ready for a variety of interesting applications in solar and stellar spectroscopy and spectropolarimetry. We plan to make it available to the astrophysical community in the near future with the hope that it will facilitate new advances in solar and stellar physics.

\begin{acknowledgements}
Financial support by the Grant Agency of the Czech Republic through grant \mbox{P209/12/P741}, by the project \mbox{RVO:67985815}, as well as by the Spanish Ministry of Economy and Competitiveness through projects \mbox{AYA2010--18029} (Solar Magnetism and Astrophysical Spectropolarimetry) and CONSOLIDER INGENIO CSD2009-00038 (Molecular Astrophysics: The Herschel and Alma Era) is gratefully acknowledged. We are also grateful to the European Union COST action MP1104 (Polarization as a Tool to Study the Solar System and Beyond) for financing short-term scientific missions that facilitated the development of this research project. The authors acknowledge the generous computing time grants by the Spanish Supercomputing Network at the LaPalma Supercomputer, managed by the Instituto de Astrof\'\i sica de Canarias, as well as by the BSC (MareNostrum, National Supercomputing Center, Barcelona). 
\end{acknowledgements}

\appendix

\section{Atomic and atmospheric models\label{app:atom}}

The numerical benchmarks shown in this paper result from multilevel radiative transfer computations in a 3D model of the solar atmosphere. The model atmosphere was obtained by imposing horizontal sinusoidal fluctuations of the kinetic temperature at each height in model C of \citet{fal93} (hereafter FAL-C model),
\begin{equation}
T(x,y,z) = T_{\rm FAL-C}(z) + \Delta T \sin\frac {2\pi x}{L_x} \sin\frac{2\pi y}{L_y}\,,
\end{equation}
where $L_x=L_y=2000$\,km are the horizontal dimensions of the domain and $\Delta T=500$\,K is the amplitude of temperature perturbation; $T_{\rm FAL-C}(z)$ corresponds to the temperature at each height $z$ in the FAL-C model. The vertical extent of the model is from $z_{\rm min}=-100$\,km to $z_{\rm max}=2\,100$\,km. 

The discretization of the spatial grid we used is $N_x\times N_y\times N_z = 100\times 100\times 121$ grid points, with periodic boundary conditions along the $x$- and $y$-axis. In order make the model more numerically demanding, we interpolated the physical quantities of the original grid of the FAL-C model onto a finer grid in the $z$ direction using an equidistant spacing of approximately 18\,km. A uniform grid spacing of 20\,km has been used for the $x$ and $y$ directions. We point out that uniform spacing is not necessary for reaching convergence with PORTA. 

The atomic model is the same five-level Ca\,{\sc ii} model described in \citet{rms10}, with the following spectral lines: the H (3969\,\AA) and K (3934\,\AA) UV lines and the infrared triplet (8498, 8542, and 8662\,\AA). The atomic level polarization created by anisotropic radiation pumping is fully taken into account, including the effects of dichroism (selective absorption of polarization components) due to the atomic polarization of the lower (metastable) levels of the infrared triplet.

The total number of discrete radiation frequencies is 505 (101 per spectral line). The angular quadrature consists of five inclination angles per octant (using Gaussian quadrature) and four azimuthal angles per octant (using the trapezoidal rule), which implies 160 ray directions.

In this paper, we illustrate the performance of PORTA through multilevel radiative transfer calculations in the 3D model of the solar atmosphere described above. We point out, however, that we have also successfully performed full multilevel computations in more complicated 3D models of the extended solar atmosphere resulting from state of the art radiation magneto-hydrodynamic simulations \citep[e.g., see][]{stepan12}.

\section{The software and hardware used for benchmarking\label{app:bench}}

The PORTA software was written in the C~programming language with parallelization treated within the Message Passing Interface (MPI) standard.\footnote{\url{http://www.mcs.anl.gov/research/projects/mpi/}} The core of PORTA consists of a shared library called PORTAL, providing the radiative transfer functionality, management of the grids, and the treatment of parallelization. The second part of PORTA is a wrapper processing, in a command-line manner, with the user inputs and outputs.

The flexibility of PORTA results from the use of additional libraries (modules) making it a very flexible code suitable for possible generalizations and/or particularizations. Different atomic models and options for applications in spectroscopy and/or spectropolarimetry can be chosen using these modules. It is also possible to choose particular or limiting cases, such as the regime of scattering polarization and the Hanle effect, or the Zeeman effect regime with or without atomic level polarization, or simply to consider the case of the non-LTE problem of the 1st kind \citep{mihalas78}. Moreover, the logical structure of PORTA is very suitable for future generalizations aimed at solving 3D problems with partial frequency redistribution. The software modules of PORTA are loaded on demand and provide a scalable functionality for further development of the code. We note that these modules can be written in various programming languages.

The numerical tests of PORTA presented in this paper were carried out in the following  computer clusters:
\begin{enumerate}
\item The Ond\v{r}ejov Cluster for Astrophysical Simulations (OCAS) operated by the Astronomical Institute of the Academy of Sciences of the Czech Republic. The cluster consists of 16 double-core, 4 quad-core and 4 oct-core nodes (i.e., 80 CPU cores in total) with 64bit AMD Opteron CPUs @2.6\,GHz interconnected via the 4X\,InfiniBand network with a bandwidth of 800\,MBytes/s and a  latency of 5\,$\rm\mu$s. PORTA was compiled from the source code using the PathScale\texttrademark\/ compiler and the parallel functionality was provided by the OpenMPI\footnote{\url{http://www.open-mpi.org}} library.
\item Additional testing of the code using a larger number of CPUs was performed at the LaPalma supercomputer operated by the Instituto de Astrof\'isica de Canarias. The cluster consists of 1024 cores of 64bit IBM PowerPC 970 processors and uses a high bandwidth Myrinet network for efficient inter-process communication. We used the GNU C compiler.
\end{enumerate}

We have also used the MareNostrum III supercomputer of the Barcelona Supercomputing Center (BSC), which with its 48896 Intel Sandy Bridge processors in 3056 nodes is at present the most powerful supercomputer in Spain and holds the 29th position in the TOP500 list of fastest supercomputers in the world. In this supercomputer we had 2048 CPU cores at our disposal and using 800, 1200, and 2000 CPU cores we tested the scalability of our formal solver of the Stokes-vector transfer equation for the resonance polarization problem of the hydrogen Ly-$\alpha$ line in a 3D model atmosphere with $N_x\times N_y\times N_z=504\times 504\times 321$ grid points. Using the Intel C/C++ 13.0 compiler, we found that PORTA scales almost linearly up to at least this number of CPU cores.

\end{document}